\documentclass[a4paper,fleqn]{cas-sc}

\usepackage{CJKutf8}

\usepackage[numbers]{natbib}
\usepackage{lscape} 
\usepackage{graphicx}
\usepackage{overpic}
\usepackage{subfigure}
\usepackage{dcolumn}
\usepackage{multirow}
\usepackage{diagbox}
\usepackage{hyperref}
\usepackage{lineno}
\usepackage{fancyhdr}
\hypersetup{
    colorlinks = true,
    urlcolor   = blue,
    linkcolor = blue,
    citecolor  = blue,
}
\def\tsc#1{\csdef{#1}{\textsc{\lowercase{#1}}\xspace}}
\tsc{WGM}
\tsc{QE}

\begin{document}
\let\WriteBookmarks\relax
\def\floatpagepagefraction{1}
\def\textpagefraction{.001}

\shorttitle{Effects of appendages on the turbulence and flow noise of a submarine model using high-order scheme}    

\shortauthors{P. Jiang et al.}  

\title [mode = title]{Effects of appendages on the turbulence and flow noise of a submarine model using high-order scheme}

\author[1]{Peng Jiang}[orcid=0000-0002-9072-8307]
\credit{Methodology, Software, Validation, Writing – original draft, Formal analysis, Visualization}

\author[1,2]{Shijun Liao}[orcid=0000-0002-2372-9502]
\credit{Conceptualization, Supervision, Project administration, Funding acquisition}

\author[3,4]{Ling Liu}[orcid=0000-0002-5146-8063]
\credit{Conceptualization, Validation, Writing – review and editing}

\author[1,2]{Bin Xie}[orcid=0000-0002-4218-2442]
\credit{Conceptualization, Supervision, Writing – review and editing, Funding acquisition}

\cortext[cor1]{Corresponding author: Dr. B. Xie (Email: xie.b.aa@sjtu.edu.cn)}

\affiliation[1]{organization={School of Ocean and Civil Engineering, Shanghai Jiao Tong University},
                city={Shanghai},
                postcode={200240}, 
                country={China}}

\affiliation[2]{organization={State Key Laboratory of Ocean Engineering, Shanghai Jiao Tong University},
                city={Shanghai},
                postcode={200240}, 
                country={China}}

\affiliation[3]{organization={China Ship Scientific Research Center},
                city={Wuxi},
                postcode={214082}, 
                country={China}}

\affiliation[4]{organization={Taihu Laboratory of Deepsea Technological Science},
                city={Wuxi},
                postcode={214082}, 
                country={China}}
\begin{abstract}
This study employs high-fidelity numerical simulations to investigate the influence of appendages on the turbulent flow dynamics and far-field acoustic radiation of the SUBOFF submarine model at a Reynolds number of $\mathrm{Re} = 1.2 \times 10^7$.
Utilizing a third-order numerical scheme combined with wall-modeled large eddy simulation (WMLES) and the Ffowcs Williams-Hawkings (FW-H) acoustic analogy, the hydrodynamic and acoustic behaviors of an appended SUBOFF configuration are compared to those of a bare hull. A computational grid of 103 million cells resolves the intricate flow interactions, while 648 hydrophones positioned 500 diameters from the model capture far-field acoustic signatures.
Key results reveal that appendages significantly amplify hydrodynamic and acoustic disturbances. Flow separations and vortex shedding at appendage junctions elevate pressure-induced drag contributions, contrasting the viscous-dominated drag of the bare hull. The sail-hull interaction intensifies local surface pressure fluctuations, increasing power spectral density (PSD) amplitudes by up to an order of magnitude. In the far field, the appended SUBOFF generates sound pressure levels approximately 20 dB higher than the bare hull, with distinct dipole directivity patterns and peak noise levels (85.10 dB) observed on the central plane. Appendages also disrupt wake symmetry, introducing complex vortical structures such as horseshoe and necklace vortices.
These findings demonstrate the critical influence of appendages on hydrodynamic and acoustic behavior, filling a gap in turbulence noise research for complex underwater geometries and providing a vital foundation for the noise reduction optimization of advanced underwater vehicles.
\end{abstract}


\maketitle

\section{Introduction}\label{introduction}

The acoustic characteristics of naval vessels, particularly underwater vehicles, have become pivotal in determining their stealth capabilities, especially in the case of submarines. The ability to operate with minimal noise has emerged as a critical aspect of operational efficacy for submarines, as evidenced by numerous studies \cite{Tang2019Noisebook,Wang2006noiseReview,Yu2007shipNoise,Li2016review,He2024Review}. \textcolor{black}{There are three main categories of submarine noise: mechanical, propeller, and hydrodynamic noise \cite{Yu2007shipNoise}. Hydrodynamic noise can be subdivided into flow-related noise and flow-induced noise. Flow noise arises primarily from turbulent boundary layer effects and flow instabilities, generating pressure fluctuations in the fluid. Flow-induced noise, on the other hand, results from surface pressure fluctuations caused by fluid-body interactions, causing excitation and vibration of the body’s structure.} While flow noise generally has a relatively low sound pressure level (\textit{SPL}) in comparison to machinery and propeller noise, its contribution becomes increasingly significant as submarine speeds rise. Machinery noise remains relatively stable across speed variations, whereas flow noise exhibits an exponential increase, often following a fifth-to-sixth power relationship, as demonstrated by \citet{Tang2019Noisebook}. Consequently, accurately predicting flow noise induced by turbulent flow around a submarine is essential for both civil and military applications, particularly for optimizing stealth performance.

Recent advancements in the study of turbulent flow around submarine models have predominantly focused on bare hull configurations. While the literature on bare hulls is extensive, only a subset of studies \cite{Chen2023SUBOFF, morse2021suboff, Kumar2018SUBOFF, LIU2023112009, Jiang2024SUBOFF, Jimenez2010, Ortiz2021slenderBody} is highlighted here, emphasizing those most pertinent to the current research. In contrast, investigations into fully appended submarine models \cite{posa2016suboff, Liu2023suboffVortex, wang2021numerical, Zhou2022suboff, Qu2021suboff} remain limited due to their geometric complexity and the intricate flow dynamics they induce. The presence of appendages significantly modifies the flow field, introducing additional vortical structures and altering turbulence characteristics. For example, the interaction of turbulent flow with submarine appendages generates complex vortex dynamics, including the evolution of the horseshoe vortices, the shedding of hairpin vortices from the fairwater, and the generation of vortex structures resembling necklaces around the fins.
Numerous studies have underscored the critical role of appendages in reshaping flow characteristics and turbulence structures. For instance, Huang et al. \cite{Huang1992Exp} conducted experimental analyses of the turbulent flow behind the SUBOFF models with various appendages as well as the SUBOFF hull, demonstrating that appendages significantly influence velocity and pressure distributions at the propeller disc position. Similarly, Liu et al. \cite{Liu2011} reduced the strength of the horseshoe-like vortex by applying a fairing, which enhanced the consistency of the flow at the propeller plane. Furthermore, Holloway et al. \cite{Holloway2015} have comprehensively analyzed flow separation phenomena induced by appendages on the SUBOFF model, highlighting the complexity of these interactions. Additionally, Leong et al. \cite{Leong2016} examined flow field variations across three configurations of the Joubert submarine during steady turning maneuvers. Their findings revealed that the bare hull generates horizontally symmetric vortices. In contrast, the sail and sail fairing respectively reduce and increase the size of vortices in the upper part of the model. Notably, the vortex formed at the tip of the sail fairing is considerably larger compared to those observed in the lower portion of the vessel. Collectively, these studies demonstrate that appendages play a pivotal role in modifying wake characteristics and turbulence structures, which directly influence submarine noise generation.

Despite these insights, a significant gap remains in understanding how appendages influence wall pressure fluctuations and turbulent noise in fully appended submarine models. Given that wall-pressure fluctuations are primarily related to the generation of the flow noise, elucidating the relationship between appendage-generated vortical structures and surface pressure spectra is critical for assessing submarine acoustic stealth. Compared to a bare hull, an appended submarine introduces multiple localized flow interactions, such as vortex impingement, pressure fluctuations, and unsteady shear layers, which may serve as dominant acoustic sources. For example, \citet{Jimenez2010} demonstrated that a supporting NACA airfoil, similar in configuration to a submarine sail, can significantly alter local flow characteristics. Furthermore, \citet{Jimenez2010Effects} found that stern fins disrupt the self-similarity of wake flow and enhance turbulence intensity and shear stress at fin tip locations, reinforcing the idea that appendages play a substantial role in near-wall pressure disturbances. However, systematic investigations into the effects of appendages on turbulent pressure fluctuations and noise generation remain limited.
Recent studies have begun to address this gap. For instance, \citet{wang2021numerical} investigated how the shape of the leading edge of the fairwater affects submarine flow noise, reporting that sails generate significant flow noise and that optimizing sail shape can effectively reduce sound levels. Rocca et al. \cite{Rocca2022BB2} examined flow past the BB2 submarine model at $\mathrm{Re} = 1.2 \times 10^6$, focusing on the near-field flow noise. The results showed that significant flow separation in the wake is the main source of near-field noise. This noise is characterized by a low-frequency broadband spectrum in the sound pressure level. Similarly, \citet{Ma2024OE} analyzed the influences of vortex shedding and turbulence on turbulent noise, concluding that localized flow interactions at appendages significantly influence far-field acoustic signatures. Additionally, \citet{Ma2025JFS} studies the effects of serrations, which are periodic notches or ridges placed along the edges of a submarine sail, on the flow and noise characteristics. Their study revealed that serrations effectively reduce both dipole and quadrupole noise sources, achieving a total noise reduction of up to 4.32 dB.
While these studies have significantly advanced the understanding of turbulent noise in submarines, several important aspects remain unresolved. These include the far-field sound directivity across all spatial directions, the characteristics of surface pressure fluctuations, and the effects of appendages on turbulence and noise generation. Therefore, further research is important to systematically study the influence of appendages on the wall pressure fluctuations and turbulent noise characteristics of fully appended axisymmetric bodies.

Despite extensive advancements in turbulent flows around submarines, the study of flow noise for fully appended submarines using high-fidelity large-eddy simulation (LES) approaches remains largely unexplored. In addition, predicting flow noise is particularly challenging. This is caused by the high computational cost of simulating high Reynolds numbers ($\mathrm{Re} \sim 10^{7-8}$). For instance, direct numerical simulation (DNS) demands prohibitively large grids, scaling as $\mathcal{O}(\mathrm{Re}^{37/14})$, while wall-resolved large-eddy simulation (WRLES) reduces this requirement to $\mathcal{O}(\mathrm{Re}^{13/7})$. However, even with this reduction, the computational cost remains prohibitively high \cite{Choi2012gridWMLES}. As a more feasible alternative, wall-modeled large-eddy simulation has been employed to bridge the gap between near-wall flow and outer flow physics, significantly reducing grid requirements to $\mathcal{O}(\mathrm{Re})$. WMLES has been successfully applied in several studies, including the SUBOFF model, yielding results that align well with experimental data \cite{Jiang2024SUBOFF}.
Moreover, traditional second-order numerical schemes often prove inadequate for vortex-dominant flows and computational acoustics due to excessive numerical dissipation. In contrast, recent advancements in high-order methods, such as the Finite Volume method based on Merged Stencil with third-order reconstruction (FVMS3) \cite{Xie2019High-fidelitysolver}, offer at least third-order accuracy with reduced numerical dissipation. This makes them particularly suitable for simulating high-Reynolds-number turbulent flows. The FVMS3 method has been effectively applied in marine hydrodynamics, including simulations of turbulent wake flow \cite{Jiang2024Sphere} and flow noise of the SUBOFF hull \cite{Jiang2024SUBOFF}.
Previously, our research focused on validating the high-order noise simulation model for accurately resolving complex turbulence and flow noise, albeit limited to the bare hull of the SUBOFF model. While these advancements provide a solid foundation, extending such high-fidelity approaches to fully appended submarine models remains an open and critical area for future research. 

Building on the aforementioned background, this paper extends the validated model to explore turbulent flow and far-field sound fields of the fully appended SUBOFF model, which includes a fairwater and four fins. Moreover, the aim of this research is to investigate the effects of appendages on turbulence and flow noise by comparing them with previous results from the bare SUBOFF hull. Specifically, the study attempts to address the following problems: (i) What are the characteristics of hydrodynamic forces and turbulent flow, particularly in terms of how appendages affect drag forces, velocity distributions, and vortex formation? (ii) What is the distribution of surface pressure fluctuations and how do they influence noise generation, particularly in relation to appendage-induced vortex dynamics and pressure variations along the hull?(iii) How do the appendages (sail and fins) affect the far-field acoustic distribution, especially in terms of \textit{SPL} differences, directivity patterns, and their relationship with turbulent structures of different scales induced by the appendages? In order to address these questions comprehensively, this study employs the WMLES method for simulating turbulent flow around the appended SUBOFF model and the acoustic analogy to calculate far-field flow noise at a Reynolds number of \(\mathrm{Re} = 1.2 \times 10^7\). Furthermore, a refined mesh comprising 103 million control volumes is used, and 648 hydrophones positioned 500 diameters from the model facilitated an extensive simulation, which can offer valuable insights into the distribution of acoustic pressure across all spatial directions.

The structure of this paper is outlined as follows: the numerical framework is presented in Section~\ref{methodology}, including the solution of fluid dynamics and the non-equilibrium wall model. Subsequently, Section~\ref{suboff} provides detailed information on the numerical setup, including the flow problem, the resolution of the numerical simulations, and specifics regarding the acoustic post-processing. Then, Section~\ref{results} presents a detailed discussion of the effects of appendages on the turbulence and flow noise of the appended axisymmetric body. Finally, the paper concludes with a summary of key findings in Section~\ref{conclusions}.

\section{Methodology}\label{methodology}
\subsection{Governing equations}\label{governing_equations}
The governing equations are given as the spatially filtered incompressible Navier–Stokes equations, which can be expressed by the Einstein notation as follows:
\begin{gather}
    \frac{\partial \widetilde{u}_i}{\partial x_i}=0, \label{eq-cont}
\\[3pt]
    \frac{\partial \widetilde{u}_i}{\partial t}+\frac{\partial \widetilde{u}_i \widetilde{u}_j}{\partial x_j}=-\frac{\partial \widetilde{p}}{\partial x_i}-\frac{\partial \tau_{i j}}{\partial x_j}+\nu \frac{\partial^2 \widetilde{u}_i}{\partial x_j \partial x_j},\label{eq-momnt1}
\end{gather}
where $i$ and $j$ are indices corresponding to the three spatial directions and the tilde symbol $\widetilde{\cdot}$ indicates spatial filtering over the grid. Note that the coordinates along the streamwise, wall-normal, and spanwise directions are denoted as $x_1$, $x_2$, and $x_3$ respectively. Moreover, $\widetilde{u}_i$ represents the filtered velocity component,  $\widetilde{p}$ is the filtered pressure divided by density, and $\nu$ denotes the kinematic viscosity, $\tau_{ij}$ represents the SGS stress tensor. 
In this study, the wall-adaptive local eddy-viscosity (WALE)~\cite{Nicoud1999WALE} model is used for the LES closure.

To enable high-fidelity simulations, a novel numerical model is developed for solving governing equations on unstructured grids. In this framework, the spatial discretization of the momentum equations is realized by the third-order accurate FVMS3 scheme \cite{Xie2019High-fidelitysolver} which uses a merged compact stencil to get rid of the difficulty in selecting admissible cells in the reconstruction process. For pressure-velocity coupling, the fractional step approach \cite{Kim1985Application,Chorin1968Numerical} is adopted, while time advancement is handled using a third-order Total Variation Diminishing (TVD) Runge-Kutta scheme \cite{Gottlieb1998TVD}. 
The resulting numerical scheme significantly improves the numerical accuracy and dissipation and effectively reduces the dependence of numerical solutions on the grid quality, which makes it well-suited for high-Reynolds-number flows in complex geometries. This is because the conventional second-order finite volume schemes have excessive numerical dissipation that may potentially overwhelm the impact of the turbulent viscosity model within the context of LES. As demonstrated by Jiang et al. \cite{Jiang2024SUBOFF}, the proposed scheme shows significant advantages in predicting turbulent flow around the blunt body including a circular cylinder at $\mathrm{Re}=1.0 \times 10^4$ and a SUBOFF hull at $\mathrm{Re}=1.2 \times 10^7$ compared to the results of traditional schemes. Note that the present solver is developed by integrating the high-order discretization with the templates provided by the OpenFOAM to deal with unstructured grid and parallel computation. For further algorithmic details, interested readers are referred to the works of Xie et al. \cite{Xie2019High-fidelitysolver,Xie2020consistent,Jiang2024Sphere}.

\subsection{Non-equilibrium wall-model}\label{wall_model}
Wall-modeled LES is employed in our study to alleviate the stringent grid-resolution requirements in the near-wall region of wall-bounded turbulent flows. By modeling the near-wall physics through a wall model, WMLES requires grid spacings that scale with the local boundary-layer thickness rather than the viscous sublayer scale \cite{Choi2012gridWMLES}, thereby reducing the overall computational cost to scale nearly linearly with the Reynolds number. This approach allows for an orders-of-magnitude reduction in the number of grid points compared to wall-resolved LES, enabling higher-Reynolds-number simulations at a practical computational expense.

The specific wall models used in this work fall under the category of wall-stress models. These models compensate for under-resolved near-wall turbulent structures by accurately predicting and enforcing the local wall shear stress. A detailed derivation and validation of our particular implementations are provided in \citet{Jiang2024SUBOFF}; here, we give a concise overview of the non-equilibrium wall-stress model.

Unlike purely equilibrium wall models, which assume a local balance between turbulence production and dissipation, the present non-equilibrium wall model includes the effects of pressure gradients in the near-wall region. To achieve this, the boundary-layer differential equations are simplified and solved in a one-dimensional sense, sampling source terms from the resolved LES field in the outer flow. Specifically, at each time step, the wall shear stress vector is obtained by using time-averaged velocity and pressure gradients from the LES at the matching height $x_2 = h_{wm}$, measured perpendicular to the boundary surface. The governing equation for the wall-normal momentum balance in the near-wall region can be expressed as:
\begin{equation}
    \frac{\partial}{\partial x_2}\Bigl[\bigl(\nu+\nu_t\bigr) \frac{\partial\langle u_i\rangle}{\partial x_2}\Bigr]
    \;=\;\frac{1}{\rho}\,\frac{\partial\langle p\rangle}{\partial x_i}, 
    \quad (i=1,3),
    \label{eq:TBLE}
\end{equation}
where $i=1,3$ corresponds to the two wall-parallel directions, $\langle u_i\rangle$ and $\langle p\rangle$ are time-averaged velocity and pressure, respectively, and $\nu$ and $\nu_t$ are the kinematic and turbulent (eddy) viscosities. Solving this equation from $x_2=0$ (the wall) up to $x_2=h_{wm}$ provides the shear stress $\langle\tau_{w,i}\rangle$ needed by the LES to account for near-wall turbulence.

To close Eq.~(\ref{eq:TBLE}), the eddy viscosity $\nu_t$ is modeled according to \citet{Duprat2011nutmodel}:
\begin{equation}
    \nu_t
    \;=\;\nu\,\kappa\,x_2^*
      \bigl[\alpha + x_2^*(1-\alpha)^{3/2}\bigr]^\beta
      \,\Bigl[1 - \exp\Bigl(-\frac{x_2^*}{1 + A\,\alpha^3}\Bigr)\Bigr]^2,
    \label{eq:nut_model}
\end{equation}
where $A=17$ and $\beta=0.78$ are model constants. The velocity scale $u_{\tau p} = \sqrt{u_\tau^2 + u_p^2}$, introduced by \citet{Manhart2008nutmodel}, accounts for non-equilibrium effects through
\[
  u_p \;=\; \biggl|\frac{v}{\rho}\,\frac{\partial p}{\partial x}\biggr|^{1/3},
\]
where $x$ is the streamwise coordinate. The dimensionless wall-normal distance is given by $x_2^* = (x_2\,u_{\tau p})/\nu$, and $\alpha = u_\tau^2/u_{\tau p}^2$. Here, $u_\tau$ is the friction velocity and $v$ is the kinematic viscosity.

In our OpenFOAM implementation, the effective eddy viscosity $\nu_{\mathrm{eff}}$ within the LES is updated based on the wall shear stress predicted by the wall model. Specifically,
\begin{equation}
    \langle\tau_{w, i}\rangle \;=\;
    \nu_{\mathrm{eff}}\;\frac{\partial u_i}{\partial x_2},\quad (i=1,3),
    \label{eq:WMLES-nut}
\end{equation}
where $\langle\tau_{w, i}\rangle$ is computed by the wall model at $x_2 = h_{wm}$. Notably, the wall boundary condition for the velocity remains the usual no-slip condition. Instead, WMLES imposes the correct near-wall shear through adjustments to the effective viscosity boundary condition. In this way, the wall model provides a more accurate wall stress estimate by considering the pressure gradients in flow, thereby improving the fidelity of large-eddy simulations without incurring the prohibitive cost of fully resolving near-wall turbulence.

\subsection{Flow Noise Analysis}\label{Hydroacoustic_analysis}

In this work, we employ the Ffowcs Williams–Hawkings equation \cite{FWH1969} to predict far-field, flow noise in a low-Mach-number regime. Specifically, we focus on dipole-dominated noise sources arising from surface pressure fluctuations, which are captured by wall-modeled LES. These fluctuations are the primary drivers of hydrodynamic noise generation under the present low-Mach-number conditions.

It is well known that the FW-H equation can be scaled to highlight the three classical acoustic source types: monopole, dipole, and quadrupole. For compact sources \cite{Cianferra2021}, the following scaling relationships hold: 
\[
\frac{p'}{\rho_0 U_0^2}
\;\sim\;
\begin{cases}
\dfrac{D}{r} & \text{Monopole};\\[6pt]
\left(\dfrac{D}{r}\right)\mathrm{Ma} & \text{Dipole};\\[6pt]
\left(\dfrac{D}{r}\right)\mathrm{Ma}^2 & \text{Quadrupole}.
\end{cases}
\]
Here, \( p^{\prime} \) is the pressure disturbance, \( \rho_0 \) is the fluid density, \( U_0 \) is the characteristic flow velocity, \( D \) is a characteristic length scale, \( r \) is the distance between the source and the observer, and \( \mathrm{Ma} \) is the Mach number. These relationships illustrate how each source type depends on a distinct power of the Mach number.

In the present study, the flow is characterized by stationary solid surfaces, which inherently precludes any volumetric (monopole) noise sources. Thus, the monopole contribution vanishes. Furthermore, under far-field conditions, where \( D/r \sim 10^{-2} \), and with a very low Mach number (\( \mathrm{Ma} = 0.0018 \)), the dipole term (scaling with \( \mathrm{Ma} \)) overwhelmingly dominates the acoustic radiation. In contrast, the quadrupole term scales with \( \mathrm{Ma}^2 \); given that \( \mathrm{Ma}^2 \) is on the order of \(10^{-6} \), its contribution becomes negligible in comparison. Therefore, only the dipole term is computed in this analysis. This approach has been successfully applied to analogous configurations, such as far-field noise predictions for marine propellers \cite{Posa2023propeller}.

For completeness, the key terms of the FW-H equation relevant to this work can be written as:
\begin{equation}
4 \pi \hat{p}(\boldsymbol{x}, t) 
= \frac{1}{c} \frac{\partial}{\partial t} \int_{\mathbb{S}} \left[ \frac{p^{\prime} \,\widehat{n}_i \,\widehat{r}_i}{r \,\bigl| 1 - \mathbb{M}_r \bigr|} \right]_{\mathbb{T}} dS 
+ \int_{\mathbb{S}} \left[ \frac{p^{\prime} \,\widehat{n}_i \,\widehat{r}_i}{r^2 \,\bigl| 1 - \mathbb{M}_r \bigr|} \right]_{\mathbb{T}} dS.
\end{equation}

In this equation, \( \hat{p}(\boldsymbol{x}, t) \) is the acoustic pressure at the observer position \( \boldsymbol{x} \) and time \( t \). The integration surface \( \mathbb{S} \) corresponds to the solid boundary (the SUBOFF model in this case). The vector \( \boldsymbol{r} = \boldsymbol{x} - \boldsymbol{y} \) points from a source element at \( \boldsymbol{y} \) to the observer, with \( r \) as its magnitude. The unit vectors \( \widehat{r}_i \) and \( \widehat{n}_i \) denote the \( i \)-th components of \( \boldsymbol{r} \) and the surface normal, respectively. \( \mathbb{M}_r \) is the Mach number in the \( \boldsymbol{r} \) direction, assumed to be small in this study. The variables \( p^* \) and \( \rho^* \) are the free-stream reference pressure and density. The emission time \( \mathbb{T} \) can be approximated by \( t \) if \( \mathbb{M}_r \ll 1 \), indicating that propagation delays are negligible.

Throughout this work, the \textit{SPLs} are expressed in decibels (dB) referenced to \( 1\,\mu\mathrm{Pa} \). For underwater acoustics, we adopt the following reference values: the fluid density \( \rho = 1000\,\mathrm{kg/m^3} \), speed of sound \( c = 1482\,\mathrm{m/s} \), and reference pressure \( \widehat{p}_{0} = 1\,\mu\mathrm{Pa} \).

\color{black}
\section{Simulation of the fluid dynamics}\label{suboff}

\begin{figure*}[!htbp]
	\centering
	\begin{minipage}[c]{1\linewidth}
    	\centering
		\begin{overpic}[trim=0cm 6cm 0cm 4.5cm, clip, width=14cm]{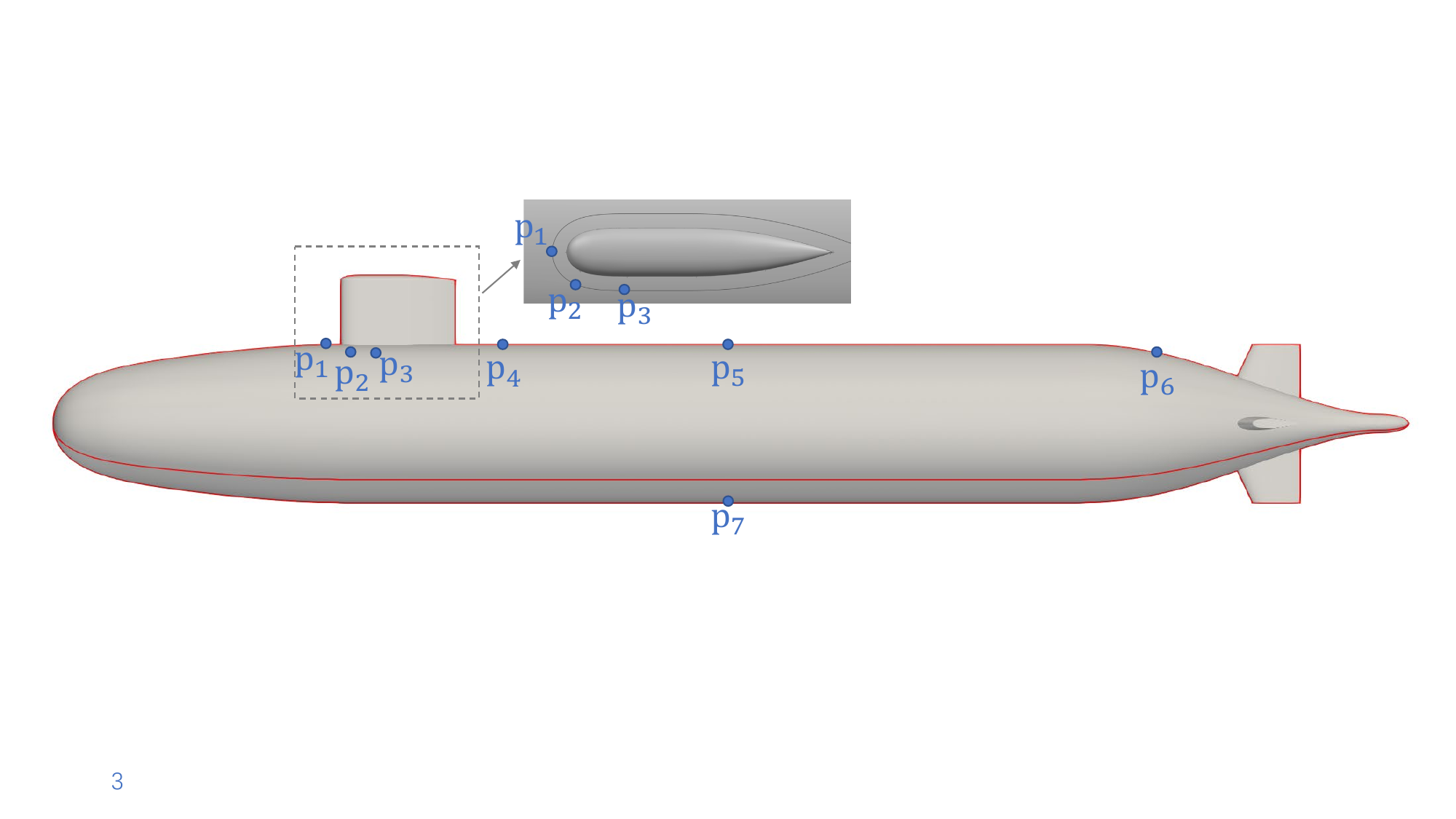}
			\put(0,20){\color{black}{(a)}}
            \put(60,16){\color{red}{Top meridian slice}}
            \put(60,7){\color{red}{Side slice}}
            \put(60,2){\color{red}{Bottom meridian slice}}
            \put(25,8){\color{black}{Hull}}
            \put(25,16){\color{black}{Sail}}
            \put(85,11){\color{black}{Fins}}
		\end{overpic}
	\end{minipage}
	\begin{minipage}[c]{1\linewidth}
    	\centering
		\begin{overpic}[trim=0cm 4cm 0.5cm 4cm, clip, width=15cm]{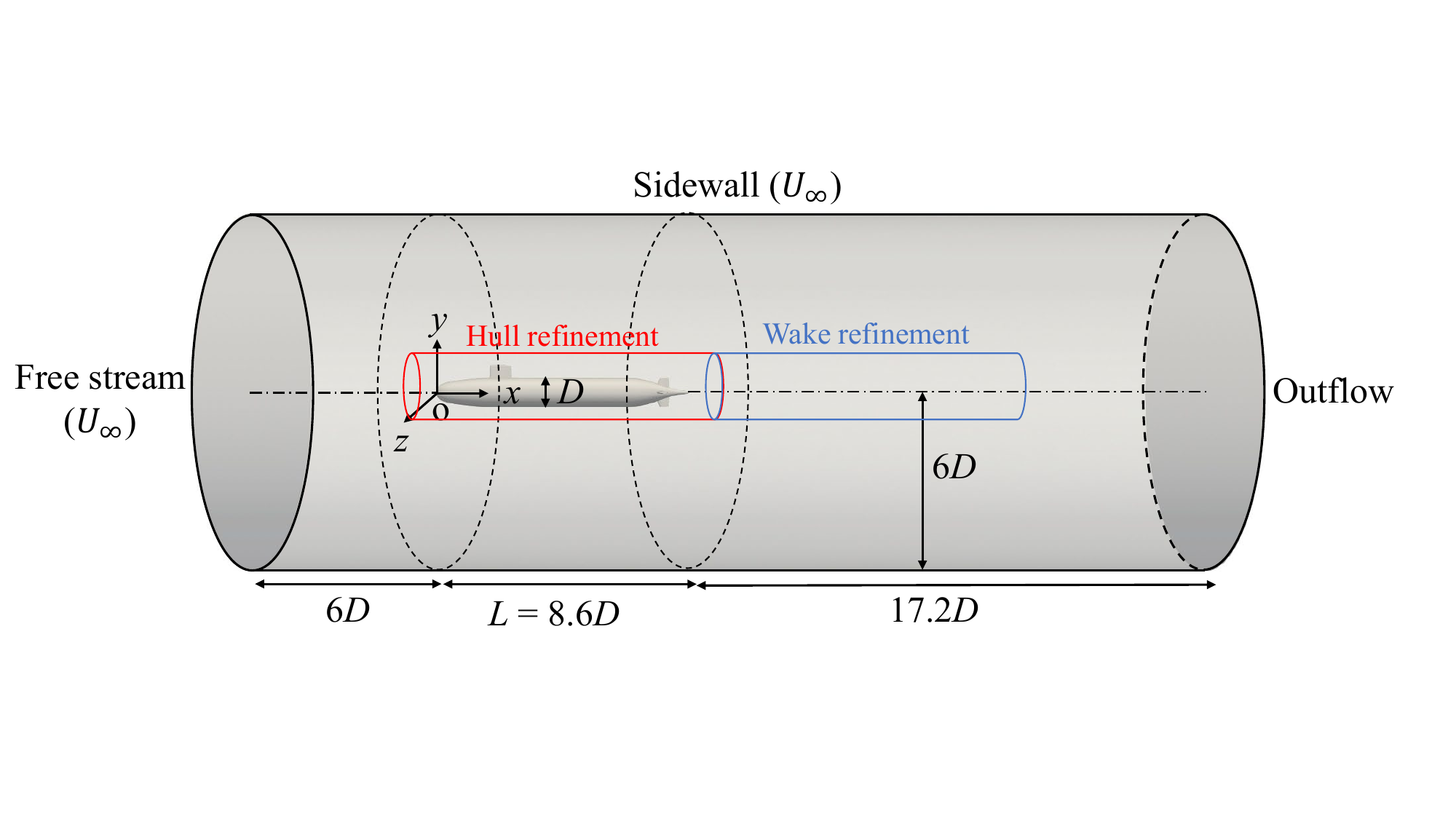}
			\put(7,30){\color{black}{(b)}}
		\end{overpic}
	\end{minipage}
	\caption[]{(a) Numerical configuration of geometry and probes for the appended DARPA SUBOFF submarine model; (b) computational set-up and boundary conditions for simulation of the appended SUBOFF model.}
	\label{fig:suboffGeo}
\end{figure*}
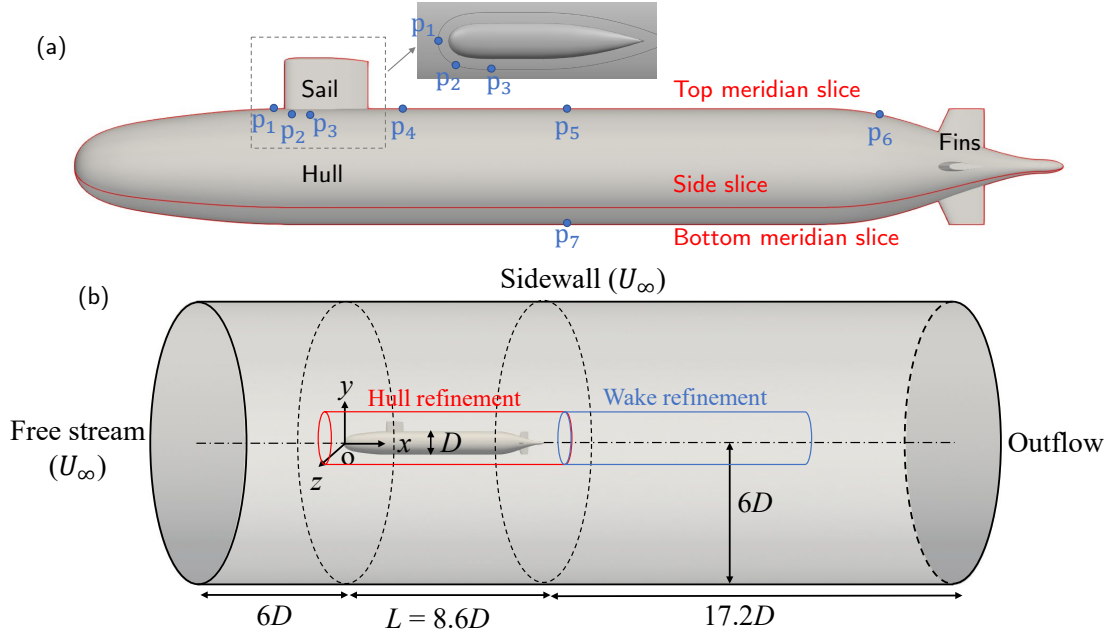

\begin{table*}[width=0.8\linewidth,cols=4,pos=htbp]
\caption{Main geometrical parameters of the DARPA Suboff (DSub) with appendages and the bare hull.}
\label{tab:suboff_parameters}
\begin{tabular*}{\tblwidth}{@{} LCCC@{} }
\toprule
\multicolumn{2}{c}{Generic submarine type}  & DSub with appendages   & DSub bare hull  \\ 
Description                    & Symbol                    & Magnitude                    & Magnitude  \\
\midrule
Length overall                    & $L$                    & 4.356 $\mathrm{m}$                    & 4.356 $\mathrm{m}$                    \\
Maximum hull diameter               & $D$                   & 0.254 $\mathrm{m}$                    & 0.254 $\mathrm{m}$                    \\
Volume of displacement            & $\Delta$                             & 0.718 $\mathrm{m^3}$ & 0.699 $\mathrm{m^3}$                    \\
Wetted surface area               & $S$                    & 6.338 $\mathrm{m^2}$ & 5.988 $\mathrm{m^2}$                  \\
\bottomrule
\end{tabular*}
\end{table*}

In this research, we utilize the DARPA SUBOFF model (DSub), a standard benchmark in submarine hydrodynamics, equipped with a sail and four stern appendages \cite{groves1989geometric}. The sail, aligned longitudinally along the upper hull, incorporates a streamlined NACA profile to reduce flow separation, while the four symmetrically positioned fins-comprising dorsal, ventral, and port/starboard stabilizers-are located at the stern to replicate realistic appendage interactions. \textcolor{black}{The simulation for the appended SUBOFF model is performed at $\mathrm{Re}_L = U_{\infty}L/\nu=1.2\times10^7$, the Reynolds number based on the inflow velocity $U_{\infty}$, the hull length of the model $L$, and the fluid kinematic viscosity $\nu$.} Fig.~\ref{fig:suboffGeo}(a) illustrates the geometric configuration, highlighting the spatial arrangement of appendages relative to the axisymmetric hull, including critical junctions such as the sail-hull and fin-hull interfaces that drive localized turbulence and acoustic sources. \textcolor{black}{For comparative analysis, numerical results are obtained from three distinct cross-sectional slice lines along the geometry: the top and bottom meridian lines, as well as a side slice line. The top and bottom slices are extracted to compare and study the influence and extent of the hull's impact on the circumferential fluid flow, while the side slice is used to investigate the influence range of the stern fins. The selection of the side slice is also based on methodologies adopted in previous studies \cite{Huang1992Exp}.} Additionally, probes were positioned at several key locations to record pressure variations, which are subsequently used for surface pressure analysis. \textcolor{black}{For completeness, Table~\ref{tab:suboff_parameters} provides the geometrical parameters that describe the primary characteristics of the DSub for both configurations-with appendages and as a bare hull.} Note that the configuration of the computational domain and boundary conditions are identical to our previous case of the SUBOFF hull in Ref. \cite{Jiang2024SUBOFF}. Fig.~\ref{fig:suboffGeo}(b) shows the computational domain setup for the appended SUBOFF model. The geometrical parameters of the computational domain are defined relative to the maximum diameter of the SUBOFF hull, $D$. Specifically, the domain extends $31.8D$  in the streamwise direction and spans a radius of $6D$ for radiation purposes. The model is positioned $6D$ from the inlet and $17.2D$ upstream of the outlet. No-slip wall boundary conditions are applied to the SUBOFF surface. Convective boundary conditions for the velocity are used at the outlet, while sidewalls are imposed with Neumann boundary conditions.
\begin{figure*}[!htbp]
	\centering
	\begin{minipage}[c]{1\linewidth}
    	\centering
		\begin{overpic}[width=14cm]{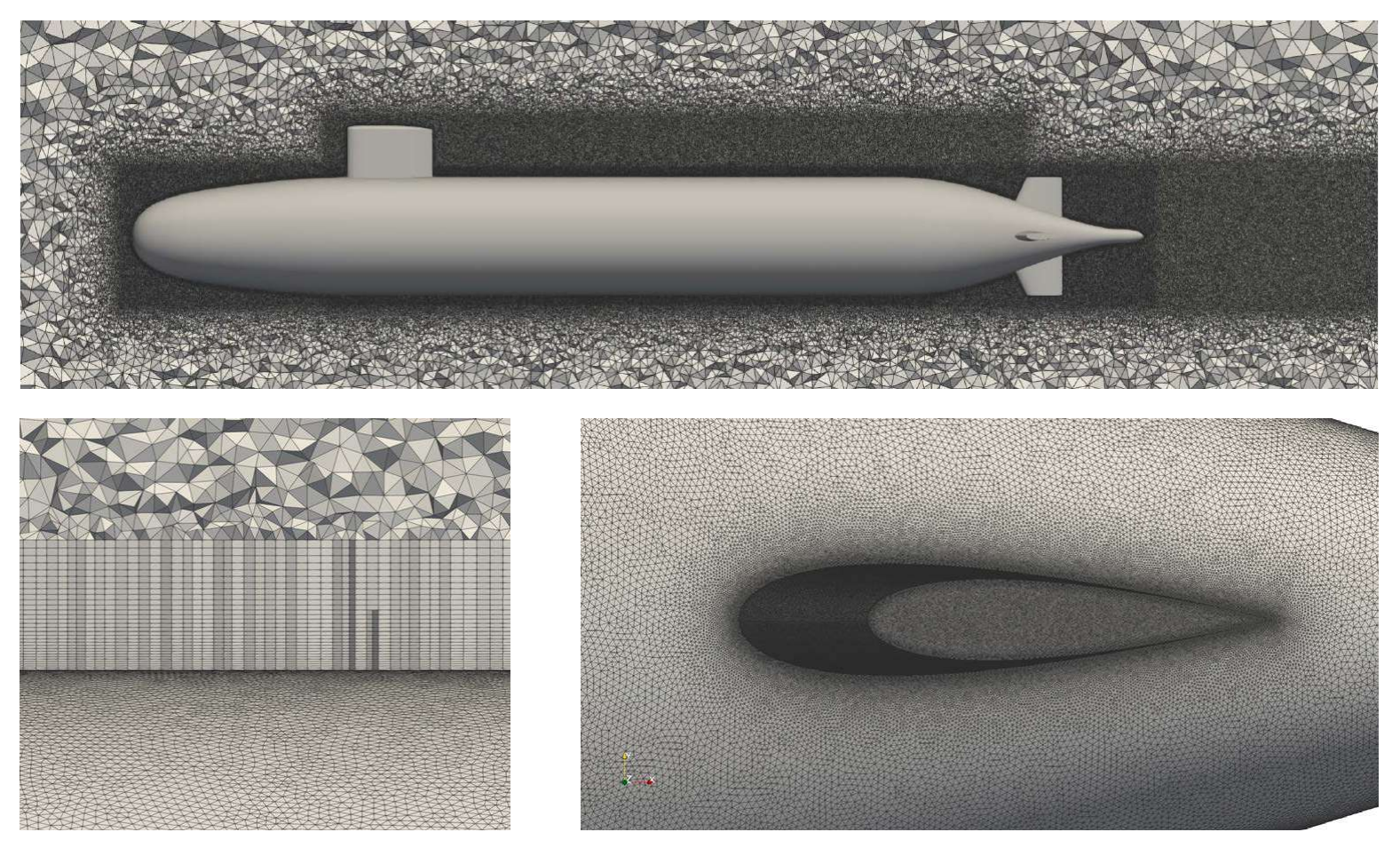}
			\put(-3,58){\color{black}{(a)}}
            \put(-3,29){\color{black}{(b)}}
            \put(37,29){\color{black}{(c)}}
		\end{overpic}
	\end{minipage}
	\caption[]{Information of the computational grid to simulate the flow around the appendaged DARPA SUBOFF model: (a) Overview of the complete mesh profile; (b) Enlarged view focusing on the boundary and surface mesh; and (c) Detailed mesh refinement near the appendages.}
	\label{fig:suboffmesh}
\end{figure*}

\textcolor{black}{One of the primary objectives of this study is to accurately calculate the turbulent noise produced by the SUBOFF model. Since fluctuating pressure near the wall is a significant source of acoustic noise, we refine the grids in both the wall-parallel and wall-normal directions to exceed the standard WMLES mesh requirements, as outlined in \citet{Park2016wallPressure}. The computational mesh consists of approximately 103 million cells. An initial wall-normal grid thickness is set to $0.02\delta$, where $\delta$ represents the boundary layer thickness. This configuration positions the first off-wall cell at a dimensionless wall distance of $\Delta y_1^+ \approx 45$, ensuring that the near-wall flow features are adequately resolved.} The mesh is nonuniform, with finer resolution near the SUBOFF wall and wake regions to ensure that the majority of generated flow structures are resolved using the minimum grid spacing. Additionally, two cylindrical regions with a diameter of $D+4\delta$ as shown in Fig.~\ref{fig:suboffGeo}(b) are applied. The region designated for ``hull refinement'' is connected to the hull and extends from a point $0.5D$ in front of the nose to a point $0.5D$ behind the end of the hull. The matching height \( h_{wm} \), where the wall models take input from the LES, is fixed at the fifth off-wall cell center, corresponding to \( h_{wm}^+ \approx 250 \) in wall units. \textcolor{black}{This methodology has been proven to perform well in simulations of the SUBOFF hull and high-Reynolds number turbulent channel flow, as demonstrated in our previous work \cite{Jiang2024SUBOFF}. Specifically, the approach has shown its reliability in accurately capturing key near-wall flow features, such as wall shear stress and turbulent fluctuations, which are essential for accurately predicting turbulent noise.} Moreover, by setting \( h_{wm} \) within the range \( h_{wm}^+ = 175 \sim 375 \), the method avoids the significant log-layer mismatch often observed when \( h_{wm} \) is placed below the buffer layer. This range has been shown to yield more accurate results, as confirmed by prior studies \cite{Hu2023WMLES, Yang2017WMLES, Owen2020WMLES}. \textcolor{black}{In addition, the ``wake refinement'' region, located behind the SUBOFF model, is also refined to capture the intricate turbulent wake structures, as shown in Fig.~\ref{fig:suboffGeo}(b). This region extends from the aft end of the model to a distance of approximately $6D$ downstream. The refinement in this region ensures that the turbulent structures in the wake are resolved with sufficient resolution, contributing to a more accurate prediction of the wake dynamics.}

To accelerate the transition to turbulence and enable comparisons with experimental results, the boundary layer is tripped at $x/D = 0.25$ from the nose. Tripwires are positioned identically to those used in previous experiments \cite{Huang1992Exp, Liu1998SUBOFFexp, Jimenez2010, Jimenez2010Effects}. Additionally, to enhance numerical efficiency, the simulation is structured into three stages as follows: 
\textcolor{black}{
(i) The initial flow development around the SUBOFF hull is simulated using the $k-\omega$ SST type RANS model at $\mathrm{Re} = 1.2 \times 10^6$. No boundary layer perturbation (trip wire) is applied during this stage. This stage is intended to develop the boundary layer and obtain a baseline flow field before transitioning to LES. The transition from this stage to the second stage is determined by monitoring the forces on the model, which must reach convergence within approximately two flow-through times. Here, flow-through time is defined as the time required for the free-stream velocity to pass through the hull length of the SUBOFF model.
(ii) The second stage begins once the model's forces have converged, and the Reynolds number is increased to $\mathrm{Re} = 1.2 \times 10^7$. In this stage, WALE-based LES without wall models is applied. This serves to simulate the flow over the hull with the boundary layer already established. The LES step ensures that turbulent features are captured accurately, and this stage prepares the flow for the subsequent WMLES stage. The transition from the second stage to the third stage occurs when the flow is sufficiently developed, as monitored by the forces and stability of the flow field, within approximately two flow-through times.
(iii) Finally, in the third stage, WMLES with the FW-H acoustic analogy is employed to simulate the turbulence and flow noise of the DSub model at $\mathrm{Re} = 1.2 \times 10^7$. At this point, the flow is fully developed, and the simulation focuses on capturing turbulent noise generated by the model. This stage collects flow quantities over two flow-through times. This methodology follows the approaches outlined in our previous study \cite{Jiang2024SUBOFF}. The transition between stages (i), (ii), and (iii) is carried out based on the flow's development, which is monitored by the forces on the model and the flow's stability, ensuring that each stage is initiated once the flow reaches the required conditions.}
In addition, the maximum CFL number is 0.7 in this study. Concerning the computation cost, 1280 processors at the High-Performance Computing (HPC) Center of Shanghai Jiao Tong University are used for the present case, utilizing approximately a million CPU-hours. Each processor is equipped with two Intel Intel Xeon ICX Platinum 8358 CPUs (2.6 GHz, 32 cores) and $16\times32$GB TruDDR4 3200 MHz RDIMM memory.

\section{Results}\label{results}
\subsection{Hydrodynamic forces and vortex structures}\label{validation}
\begin{figure}[!htbp]
	\centering
	\begin{overpic}[width=12cm]{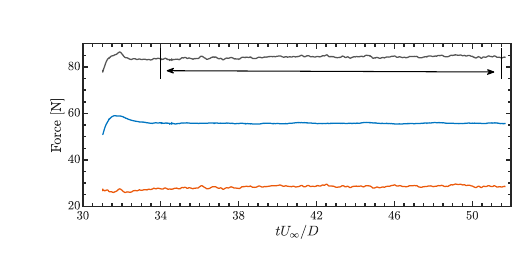}
    \put(25,36){\color{black}{$F$}}
    \put(37,32){\color{black}{two flow-through times for statistics collection}}
    \put(25,29){\color{black}{$F_\nu$}}
    \put(25,16){\color{black}{$F_p$}}
	\end{overpic}
	\caption[]{Time evolution of the drag force ($F$)  acting on the appended SUBOFF model along with the contribution from the pressure ($F_p$) and the viscous forces ($F_\nu$).}
	\label{fig:suboff_force_time_his}
\end{figure}
\begin{figure}[!htbp]
\centering
	\begin{minipage}[c]{1\linewidth}
    \centering
    \begin{overpic}[width=9cm]{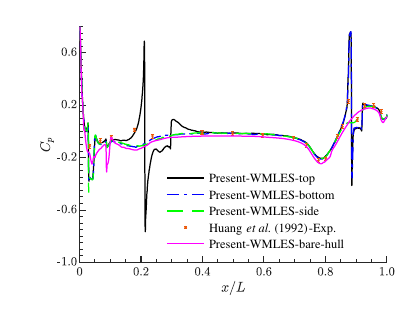}
    \put(16,-5){\includegraphics[trim=0cm 4cm 0cm 6cm, clip, scale=0.095]{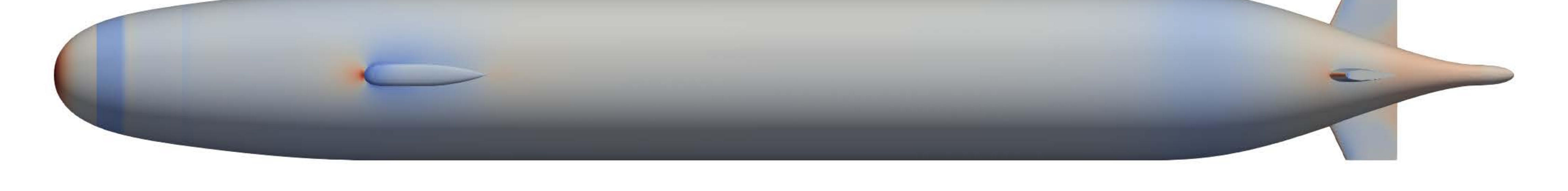}}
	\end{overpic}
	\end{minipage}
    \par\bigskip  
	\caption[]{Profiles of the mean surface pressure coefficients $C_p$ along the SUBOFF model, compared to experimental results from \citet{Huang1992Exp} and numerical results from \citet{Jiang2024SUBOFF}. The geometrical information of the top, bottom, and side slices are given in Fig.~\ref{fig:suboffGeo}(a).}
	\label{fig:suboff_Cp}
\end{figure}

\begin{figure}[!htbp]
\centering
	\begin{minipage}[c]{1\linewidth}
    \centering
	   \begin{overpic}[width=7cm]{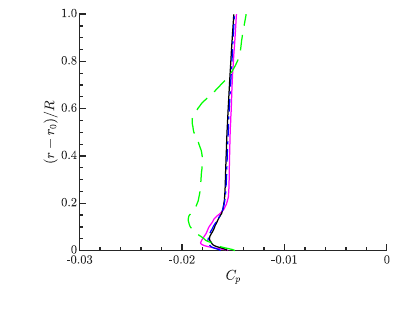}
			\put(3,70){\color{black}{\textit{(a)}}}
	   \end{overpic}
	   \begin{overpic}[width=7cm]{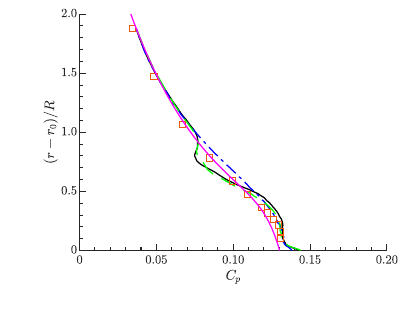}
			\put(3,70){\color{black}{\textit{(b)}}}
              \put(56,50){\includegraphics[scale=1]{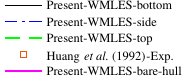}}
	   \end{overpic}
	\end{minipage}
	\begin{minipage}[c]{1\linewidth}
    \centering
	   \begin{overpic}[width=7cm]{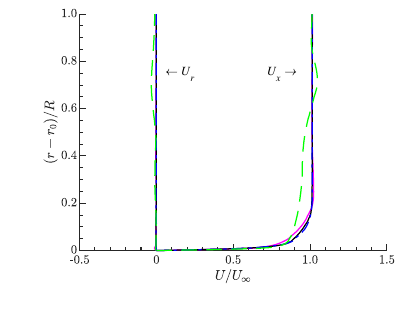}
			\put(3,70){\color{black}{\textit{(c)}}}
	   \end{overpic}
	   \begin{overpic}[width=7cm]{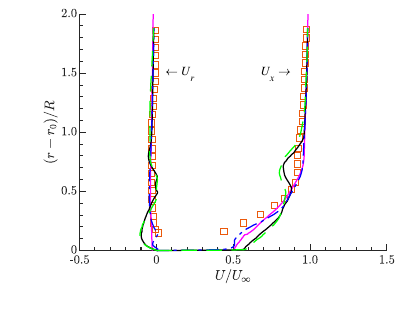}
			\put(3,70){\color{black}{\textit{(d)}}}
	   \end{overpic}
	\end{minipage}
	\caption[]{Profiles of time-averaged pressure coefficient ($C_p$) (a,b) and axial ($U_x$) and radial ($U_r$) velocity (c,d) at Slice-I: $x/L=0.502$ (a, c) and Slice-II: $x/L=0.978$ (b, d). Note that $r_0$ represents the local radius of the hull, while $R$ denotes the maximum radius.}
	\label{fig:suboff_p_U}
\end{figure}

\begin{table*}[width=0.8\linewidth,cols=5,pos=!htpb]
\caption{Comparison of drag coefficient of the appended SUBOFF model at $\mathrm{Re}=1.2\times10^7$ with available references.}
\label{tab:suboff_Cd}
\begin{tabular*}{\tblwidth}{@{} CCCCC@{} }
\toprule
\multirow{2}{*}{\textcolor{black}{Present WMLES}} & \multicolumn{2}{c}{Experimental ref.} & \multicolumn{2}{c}{Numerical ref.} \\
                    & Özden-2019 \cite{Ozden2019}         & Crook-1992 \cite{Crook1990}         & Özden-2019 \cite{Ozden2019}         & Liu et al.-2021 \cite{Liu2021URANS}          \\ \midrule
\textcolor{black}{$3.47\times 10^{-3}$}  & $3.47\times 10^{-3}$ & $3.60\times 10^{-3}$ & $3.45\times 10^{-3}$ & $3.46\times 10^{-3}$        \\
\bottomrule
\end{tabular*}
\end{table*}
To validate the numerical framework, the drag coefficient, defined as $C_D = F\left/0.5\rho U_{\infty}^{2}S\right.$ where \( F \) is the drag force and \( S \) the wetted surface area of the DARPA SUBOFF model, 
is computed and compared against prior experimental measurements and prior computational studies \cite{Ozden2019, Crook1990, Liu2021URANS}. As shown in Table~\ref{tab:suboff_Cd}, the present WMLES yields $C_D = 3.47 \times 10^{-3}$, aligning closely with experimental values ($3.47 \times 10^{-3}$ \cite{Ozden2019}, $3.60 \times 10^{-3}$ \cite{Crook1990}) and numerical simulations ($3.46 \times 10^{-3}$ \cite{Liu2021URANS}). Notably, the results exhibit less than 1\% deviation from the benchmark experimental dataset \cite{Ozden2019}, underscoring the remarkable capability of the present model in resolving hydrodynamic forces. This rigorous agreement across diverse methodologies spanning wind-tunnel tests \cite{Crook1990}, unsteady RANS \cite{Liu2021URANS}, and high-fidelity LES \cite{Ozden2019}, which confirms the robustness of the WMLES approach in capturing both viscous and pressure-induced contributions to drag, even under the complex flow separations introduced by appendages.

To systematically evaluate the impact of appendages on turbulent flow characteristics, the hydrodynamic and near-wall statistics of the appended SUBOFF model are rigorously compared to those of the bare hull. The analysis reveals significant differences in hydrodynamic force composition, mean surface pressure distribution, and near-wall velocity and pressure profiles in the stern region. As shown in Fig.~\ref{fig:suboff_force_time_his}, the temporal evolution of the drag force \( F \) is decomposed into pressure (\( F_p \)) and viscous forces (\( F_\nu \)) components, which highlights a fundamental shift in force balance. For the appended model, the viscous-to-pressure force ratio \( F_\nu / F_p \approx 1.94\) contrasts sharply with the \( F_\nu / F_p \approx 5\) reported for the bare hull in wall-resolved LES studies \citet{Kumar2018SUBOFF}. This fivefold reduction underscores how appendages amplify pressure drag through localized flow separations and vortex shedding, despite viscous forces remaining dominant overall.
The mean pressure coefficient $C_p$ along the hull is compared with reference results, which further illustrates appendage-induced effects. Along the side slice positioned between fins to minimize appendage interference, the appended model aligns closely with experimental data \citet{Huang1992Exp} and prior LES results \citet{Jiang2024SUBOFF}, validating the high fidelity of the present simulation. However, over the top and bottom meridian slices, appendages induce pronounced pressure gradients, deviating from the bare hull’s smoother distribution. These gradients correlate with regions of flow separation at the sail-hull junction and fin roots.
Near-wall velocity and pressure profiles provide additional insights into appendage-induced flow modifications. Fig.~\ref{fig:suboff_p_U} presents a comparison of the mean pressure coefficient (\( C_p \)), axial velocity (\( U_x \)), and radial velocity (\( U_r \)). At \( x/L = 0.502 \) (mid-body), the sail disrupts axial 
(\( U_x \)) and radial (\( U_r \))  velocities, elevating mid-hull pressure fluctuations. Downstream at \( x/L = 0.978 \)
(stern), fins generate asymmetric wake structures, perturbing near-wall velocity profiles and amplifying axial velocity components by up to 10\% compared to the bare hull. These disturbances propagate into the outer flow, confirming appendages’ role in altering turbulence dynamics across multiple scales. These findings highlight how appendages fundamentally alter flow physics as introducing pressure-driven drag penalties, localized flow separations, and wake asymmetries, while maintaining strong agreement with experimental data in regions where appendage-induced disturbances are minimal, such as the side slice between the fins.

\begin{figure*}[!htbp]
    \centering
    \begin{overpic}[width=13cm]{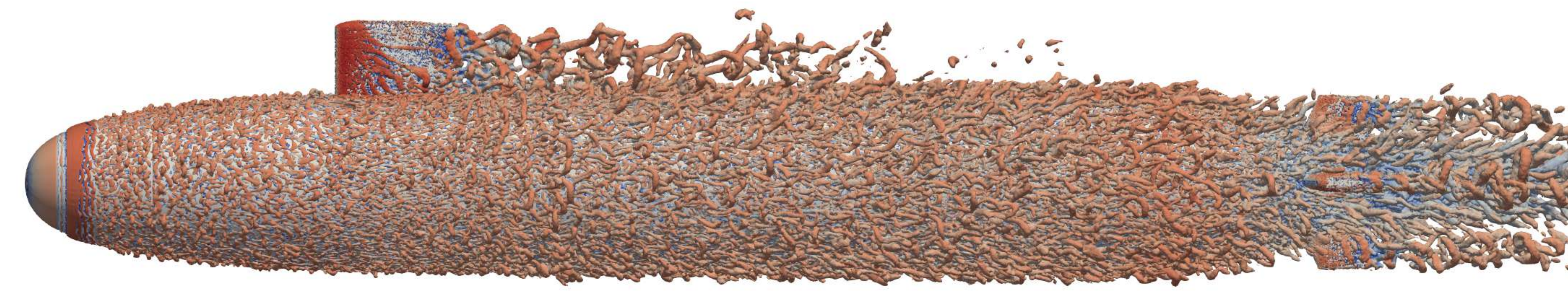}
        \put(-3,18){\color{black}{\textit{(a)}}}
        \put(80,16){\includegraphics[scale=0.15]{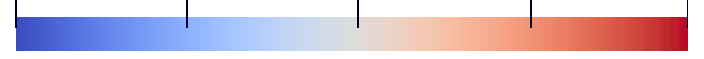}}
        \put(83,18){\footnotesize\color{black}{$u_{ins}/U_{\infty}$}}
        \put(79,18){\footnotesize\color{black}{0}}
        \put(92,18){\footnotesize\color{black}{1.41}}
    \end{overpic}
    \par\bigskip  
    \begin{overpic}[width=13cm]{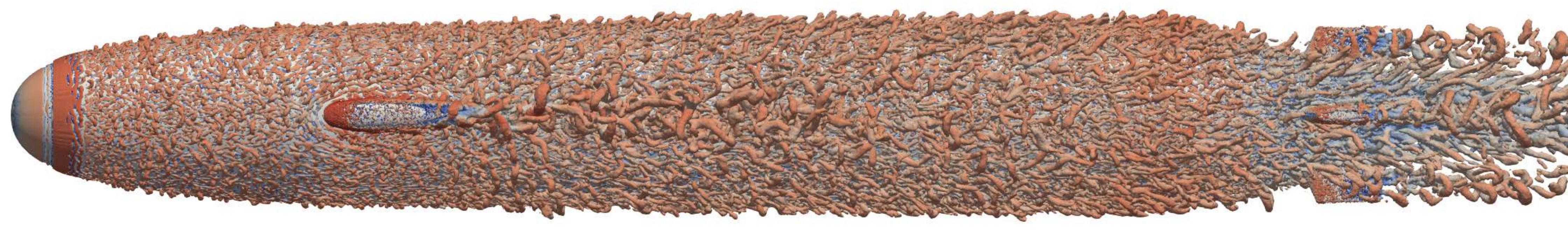}
        \put(-3,12){\color{black}{\textit{(b)}}}
    \end{overpic}
    \par\bigskip  
    \begin{overpic}[width=13cm]{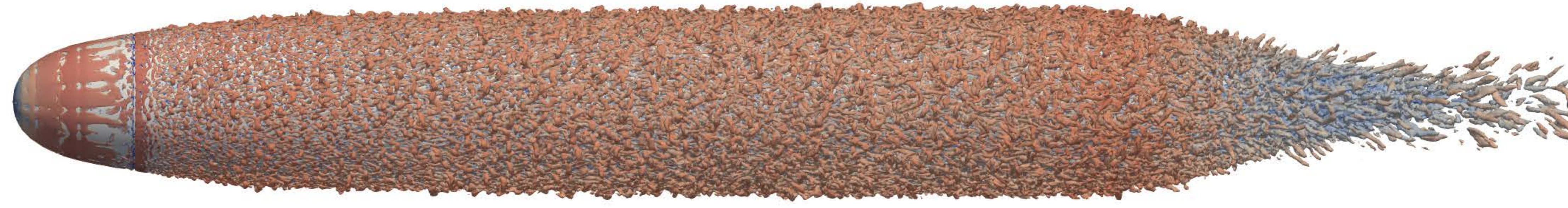}
        \put(-3,12){\color{black}{\textit{(c)}}}
    \end{overpic}
    \caption[]{Instantaneous vortex structures using the iso-surface of the $\lambda_2$ criterion with $\lambda_2=100$ at $tU_\infty/D = 50$ colored by the instantaneous velocity, (a) the front view and (b) top view of the appended axisymmetric body; (c) front view of the flow structures of the bare hull of axisymmetric body.}
    \label{fig:suboff_vortex}
\end{figure*}

\begin{figure*}[!htbp]
	\centering
	\begin{overpic}[width=13cm]{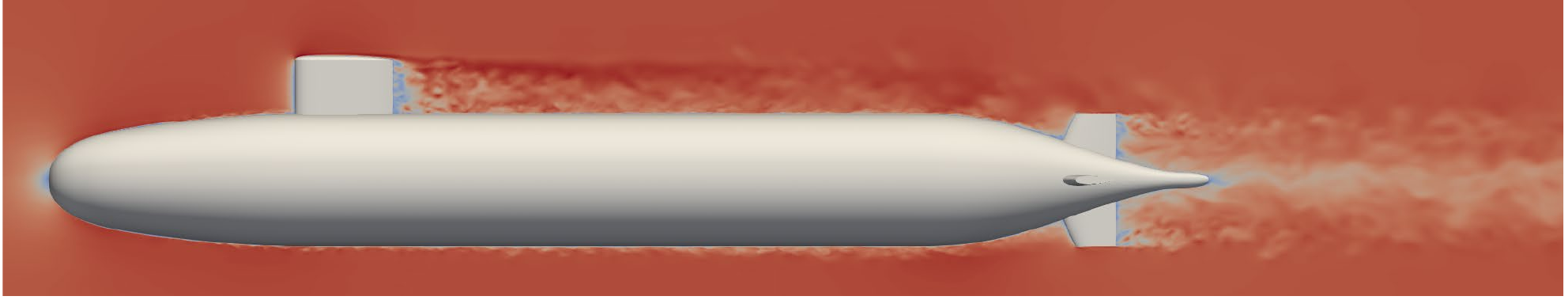}
    \put(-5,17){\color{black}{\textit{(a)}}}
	\end{overpic}
	\begin{overpic}[width=13cm]{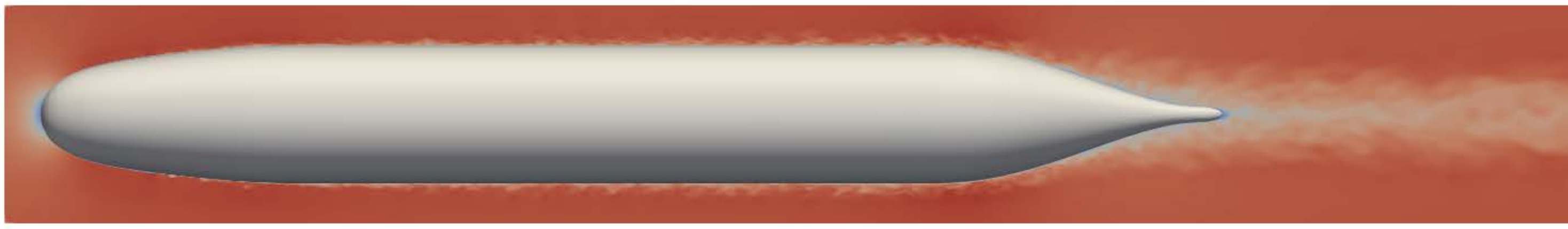}
    \put(-5,12){\color{black}{\textit{(b)}}}
	\end{overpic}
	\caption[]{Slice of the instantaneous fields of the non-dimensional streamwise velocity $u_{x, ins}/U_\infty$ ranging from -0.36 to 1.23 at $tU_\infty/D = 50$. (a) The appended SUBOFF model and (b) the SUBOFF hull.}
	\label{fig:suboff_velocity}
\end{figure*}

\begin{figure*}[!htbp]
	\centering
	\begin{overpic}[width=13cm]{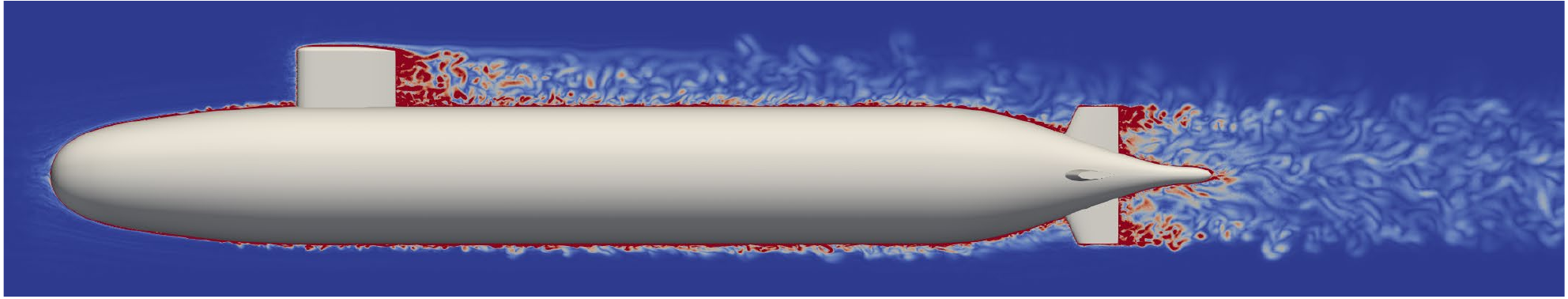}
    \put(-5,17){\color{black}{\textit{(a)}}}
	\end{overpic}
    \begin{overpic}[width=13cm]{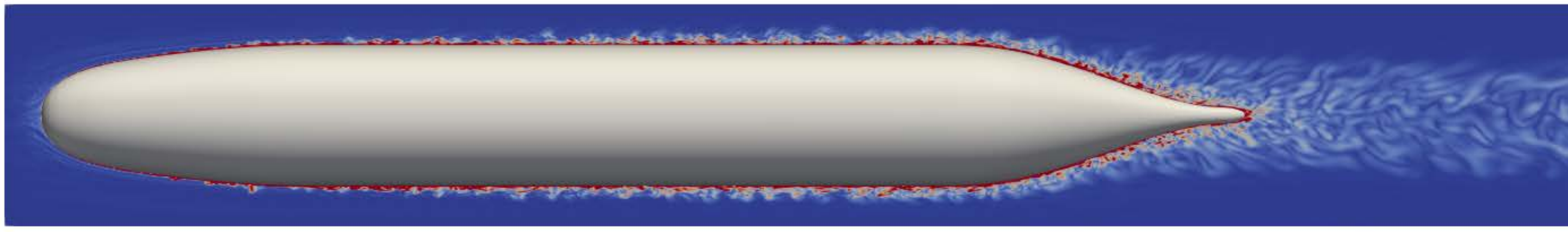}
    \put(-5,12){\color{black}{\textit{(b)}}}
	\end{overpic}
	\caption[]{(a) Slice of the instantaneous fields of the non-dimensional vorticity $\omega D/U_\infty$ ranging from 0 to 18.37 at $tU_\infty/D = 50$ for (a) the appended SUBOFF model and (b) the SUBOFF hull.}
	\label{fig:suboff_vorticity}
\end{figure*}

The turbulent coherent structures, resolved via the $\lambda_2$ criterion in Fig.~\ref{fig:suboff_vortex}, and the slice-wise distributions of streamwise velocity and vorticity in Figs.~\ref{fig:suboff_velocity} and \ref{fig:suboff_vorticity} reveal stark contrasts between the appended SUBOFF and the bare hull. 
Over the mid-hull region, both configurations exhibit common turbulent boundary layer dynamics, including near-wall streaks, stern flow separation, and periodic wake shedding driven by adverse pressure gradients, features well-documented in prior studies of axisymmetric bodies \cite{Kumar2018SUBOFF, morse2021suboff, Ortiz2021slenderBody, Jiang2024SUBOFF}. However, the appended SUBOFF introduces significant flow complexities due to interactions between the turbulent boundary layer and appendage geometries. At the sail-hull junction [Fig.~\ref{fig:suboff_vortex}(a)], a dominant horseshoe vortex forms, wrapping around the sail’s leading edge and persisting downstream with a core vorticity magnitude 1.5 higher than the bare hull’s baseline turbulence. These vortices coalesce into necklace-like structures around the fins [Fig.~\ref{fig:suboff_vortex}(c)], generating localized shear layers that obviously amplify turbulent kinetic energy compared to the bare hull.
The appendages also profoundly alter wake topology. Downstream of the fins, the appended SUBOFF exhibits a widened wake with a 35\% larger cross-sectional area than the bare hull, characterized by quasi-cylindrical vortex bundles [Fig.~\ref{fig:suboff_vorticity}(a)]. These structures induce strong velocity deficits ($U_x / U_{\infty}<0.7$) and radial velocity fluctuations ($U_r / U_{\infty} \pm 0.1$), distorting the inflow homogeneity critical for propeller efficiency. Such flow asymmetries correlate with elevated vorticity magnitudes ($\omega D / U_{\infty}>15$) near fin tips, exceeding values reported for isolated fin studies \cite{Jimenez2010Effects}. These findings align with prior investigations of appended SUBOFF models \cite{posa2016suboff, posa2020suboff}, confirming that geometric complexities amplify turbulence intensity, vortex coherence, and wake unsteadiness. The results underscore the necessity of resolving appendage-induced flow features to accurately predict hydrodynamic and acoustic performance in practical submarine designs.

\subsection{Surface pressure fluctuations}\label{Frequency _spectra}
\begin{figure}[!htbp]
	\centering
	\begin{overpic}[width=7cm]{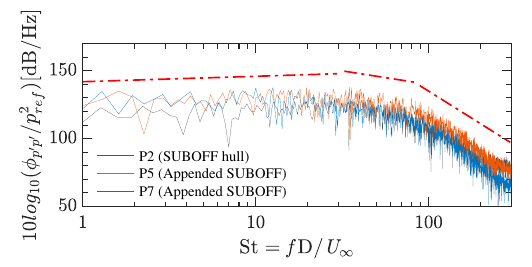}
    \put(-5,40){\color{black}{\textit{(a)}}}
            \put(30,37.5){\footnotesize{\color{red}{$\propto \mathrm{St}^{0.4}$}}}
            \put(70,37){\footnotesize{\color{red}{\rotatebox{-10}{$\propto \mathrm{St}^{-3}$}}}}
            \put(85,33){\footnotesize{\color{red}{\rotatebox{-30}{$\propto \mathrm{St}^{-9}$}}}}
	\end{overpic}
    \begin{overpic}[width=7cm]{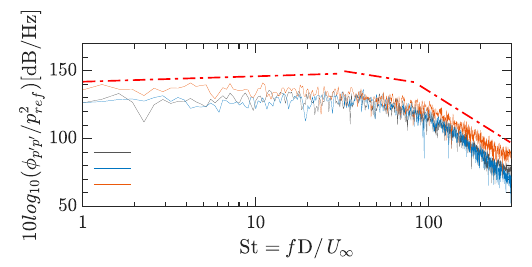}
    \put(-5,40){\color{black}{\textit{(b)}}}
            \put(30,37.5){\footnotesize{\color{red}{$\propto \mathrm{St}^{0.4}$}}}
            \put(70,37){\footnotesize{\color{red}{\rotatebox{-10}{$\propto \mathrm{St}^{-3}$}}}}
            \put(85,33){\footnotesize{\color{red}{\rotatebox{-30}{$\propto \mathrm{St}^{-9}$}}}}
            \put(25,22){\footnotesize{$\mathrm{p}_5$}}
            \put(25,18){\footnotesize{$\mathrm{p}_6$}}
            \put(25,14){\footnotesize{$\mathrm{p}_4$}}
	\end{overpic}
	\caption[]{The PSD of pressure fluctuations at various locations. (a) Point $\mathrm{p}_2$ refers to the location at $x/L = 0.501$ on the SUBOFF hull, while the other two points, $\mathrm{p}_5$ and $\mathrm{p}_7$, correspond to the top and bottom meridian planes of the appended SUBOFF model at $x/L = 0.501$, respectively. (b) Points at the top meridian planes of the appended SUBOFF model with different distances from the sail at $x/L = 0.340$ ($\mathrm{p}_4$), $x/L = 0.501$ ($\mathrm{p}_5$), and $x/L = 0.710$ ($\mathrm{p}_6$).}
	\label{fig:suboff_surface_furessure}
\end{figure}

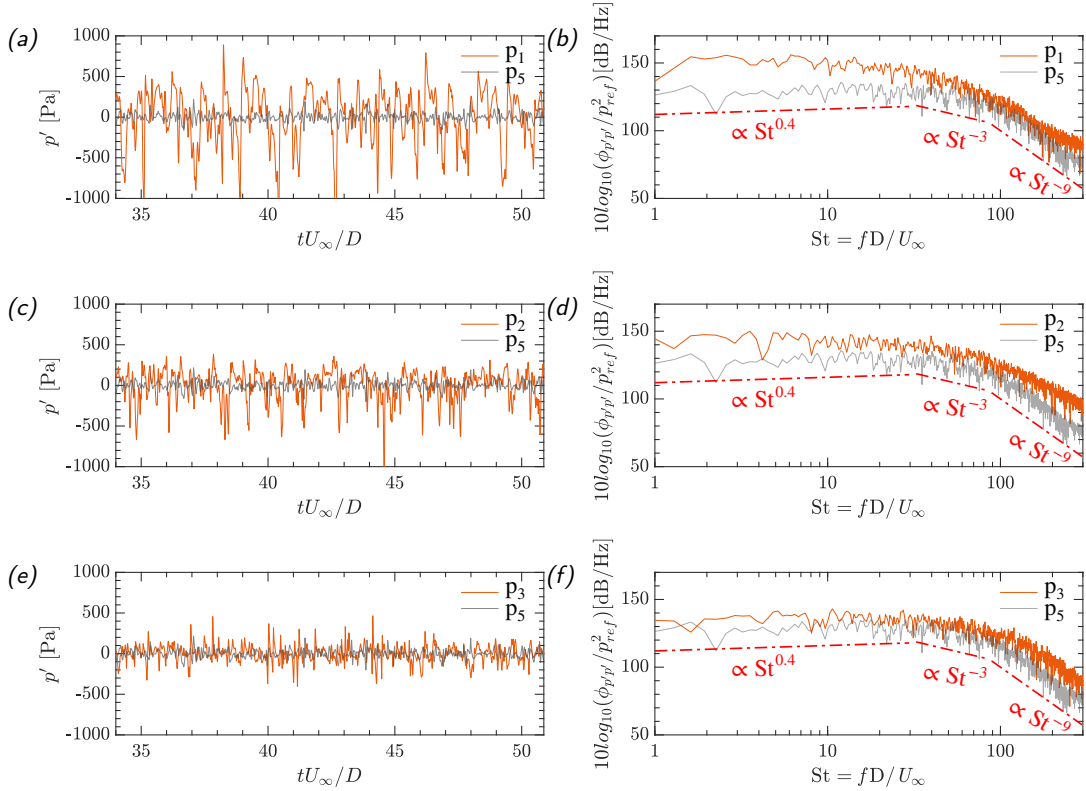
\begin{figure}[!htbp]
	\centering
    
	\begin{minipage}[c]{1\linewidth}
    \centering
	\begin{overpic}[width=7cm]{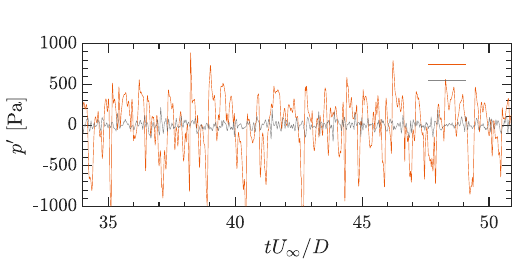}
    \put(-5,40){\color{black}{\textit{(a)}}}
    \put(89,38){\small{$\mathrm{p}_1$}}
    \put(89,34){\small{$\mathrm{p}_5$}}
	\end{overpic}
    \begin{overpic}[width=7cm]{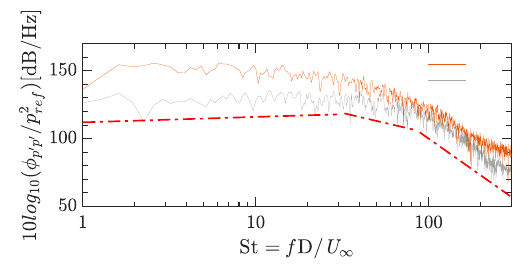}
    \put(89,38){\small{$\mathrm{p}_1$}}
    \put(89,34){\small{$\mathrm{p}_5$}}
    \put(-5,40){\color{black}{\textit{(b)}}}
            \put(30,22.5){\small{\color{red}{$\propto \mathrm{St}^{0.4}$}}}
            \put(66,22){\small{\color{red}{\rotatebox{-10}{$\propto \mathrm{St}^{-3}$}}}}
            \put(82,16){\small{\color{red}{\rotatebox{-30}{$\propto \mathrm{St}^{-9}$}}}}
	\end{overpic}
    \end{minipage}

	\begin{minipage}[c]{1\linewidth}
    \centering
	\begin{overpic}[width=7cm]{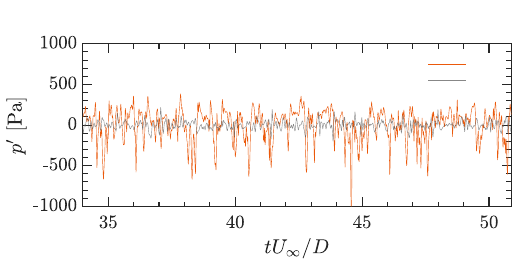}
    \put(-5,40){\color{black}{\textit{(c)}}}
    \put(89,38){\small{$\mathrm{p}_2$}}
    \put(89,34){\small{$\mathrm{p}_5$}}
	\end{overpic}
    \begin{overpic}[width=7cm]{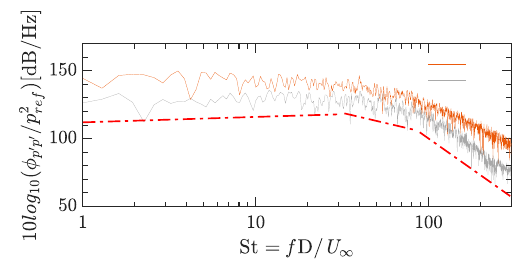}
    \put(-5,40){\color{black}{\textit{(d)}}}
    \put(89,38){\small{$\mathrm{p}_2$}}
    \put(89,34){\small{$\mathrm{p}_5$}}
            \put(30,22.5){\small{\color{red}{$\propto \mathrm{St}^{0.4}$}}}
            \put(66,22){\small{\color{red}{\rotatebox{-10}{$\propto \mathrm{St}^{-3}$}}}}
            \put(82,16){\small{\color{red}{\rotatebox{-30}{$\propto \mathrm{St}^{-9}$}}}}
	\end{overpic}
    \end{minipage}
    
	\begin{minipage}[c]{1\linewidth}
    \centering
	\begin{overpic}[width=7cm]{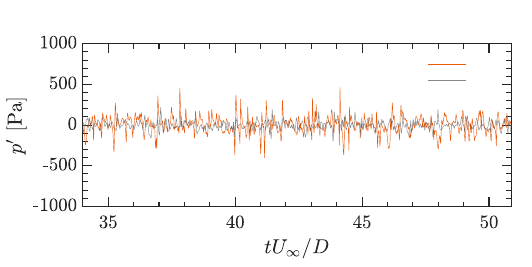}
    \put(-5,40){\color{black}{\textit{(e)}}}
    \put(89,38){\small{$\mathrm{p}_3$}}
    \put(89,34){\small{$\mathrm{p}_5$}}
	\end{overpic}
    \begin{overpic}[width=7cm]{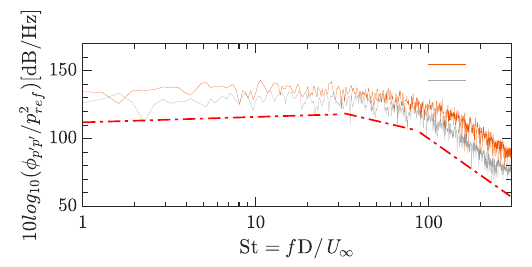}
    \put(-5,40){\color{black}{\textit{(f)}}}
    \put(89,38){\small{$\mathrm{p}_3$}}
    \put(89,34){\small{$\mathrm{p}_5$}}
            \put(30,22.5){\small{\color{red}{$\propto \mathrm{St}^{0.4}$}}}
            \put(66,22){\small{\color{red}{\rotatebox{-10}{$\propto \mathrm{St}^{-3}$}}}}
            \put(82,16){\small{\color{red}{\rotatebox{-30}{$\propto \mathrm{St}^{-9}$}}}}
	\end{overpic}
    \end{minipage}
	\caption[]{Comparisons of the time history of surface pressure fluctuations (a,c,e) and the PSD of pressure fluctuations (b,d,f) between the locations, i.e. $\mathrm{p}_1,\ \mathrm{p}_2,\ \mathrm{p}_3$ near the fairwater and mid-hull of the appended SUBOFF model $\mathrm{p}_5:x/L = 0.501$. The geometrical information of the sample locations is given in Fig.~\ref{fig:suboffGeo}(a). Locations for points are $\mathrm{p1}\ (x/L,y/L,z/L) = (0.208,0.058,0)$, $\mathrm{p2}\ (x/L,y/L,z/L) = (0.215, 0.057, -0.010)$, and $\mathrm{p3}\ (x/L,y/L,z/L) = (0.230,0.057,-0.012)$}
	\label{fig:suboff_surface_furessure_fairwater}
\end{figure}

Analyzing surface pressure fluctuations is particularly valuable for identifying sound sources and assessing the strength of turbulent flow. This section investigates the characteristics of surface pressure fluctuations, with a particular emphasis on the influence of appendages-such as the sail and fins-on the flow dynamics. The pressure probes are strategically positioned on the model, as shown in Fig.~\ref{fig:suboffGeo}(a), and categorized into three groups to facilitate a detailed comparative analysis of the time history and PSD of surface pressure fluctuations. 
Group 1: Points $\mathrm{p}_5$ and $\mathrm{p}_7$, situated at the top and bottom meridian lines, respectively, at the midpoint of the model ($x/L = 0.501$), assess the influence of the appendages on pressure fluctuations over the mid-body of the hull. 
Group 2: Points $\mathrm{p}_4$, $\mathrm{p}_5$, and $\mathrm{p}_7$, located along the top meridian line at increasing distances from the sail, are used to evaluate how the effect of the sail decreases as we move downstream along the hull. 
Group 3: Points $\mathrm{p}_1$, $\mathrm{p}_2$, and $\mathrm{p}_3$, positioned near the sail, help analyze the flow physics of pressure fluctuations in the region influenced by the complex flow patterns around the sail and fins. 
Pressure signals are recorded over two full flow-through times on the SUBOFF surface and subsequently analyzed using statistical measures and frequency-domain analyses, providing insights into the flow dynamics and acoustic implications.

Fig.~\ref{fig:suboff_surface_furessure}(a) compares the power spectral density of pressure fluctuations at point $\mathrm{p}_2$ on the SUBOFF hull and at points $\mathrm{p}_5$ and $\mathrm{p}_7$ on the appended SUBOFF model, all located at $x/L = 0.501$. The pressure levels at $\mathrm{p}_2$ are much higher (approximately a 10 dB increase) compared to those at $\mathrm{p}_5$ and $\mathrm{p}_7$, particularly in the low-frequency range ($\mathrm{St}<10$), due to the influence of the sail. This low-frequency content is associated with large-scale pressure structures in wall-bounded turbulence. The interaction of the appendages with the flow field results in the formation of distinct vortex structures, such as the horseshoe vortex around the sail and the tip vortices generated by the fins. These large-scale vortices cause considerable disturbances in the flow, leading to increased pressure fluctuations along the hull. In contrast, examining the frequency behavior at points $\mathrm{p}_5$ and $\mathrm{p}_7$ on the appended SUBOFF model, the high-frequency region at $\mathrm{p}_7$ (located on the bottom meridian slice), which corresponds to pressure structures with small wave numbers, shows a nearly 5 dB decrease compared to $\mathrm{p}_5$ (located on the top meridian slice). This indicates that the sail at the top enhances pressure structures with small wave numbers. Moreover, the similarity in the frequency behavior of the pressure fluctuations at points $\mathrm{p}_2$ (on the bare hull) and $\mathrm{p}_7$ (on the bottom meridian slice of the appended SUBOFF) indicates that the influence of the sail on the upper mid-body is strong but does not extend to the lower hull. This observation is consistent with the earlier analysis of vortex structures around the sail. These results support the observation that the presence of the appendage (sail) predominantly affects pressure fluctuations on the upper part of the hull. Fig.~\ref{fig:suboff_surface_furessure}(b) presents the power spectral densities measured at various distances from the sail along the top meridian planes of the appended SUBOFF model. At point $\mathrm{p}_4$, located $0.043L$ from the sail’s trailing edge, the PSD exhibits a significantly higher amplitude (approximately an 8 dB increase) in the low-to-mid frequency range ($1 < \mathrm{St} < 40$) than at points $\mathrm{p}_5$ and $\mathrm{p}_6$, located $0.204L$ and $0.413L$ downstream, respectively. The reduced low-frequency pressure fluctuations at further downstream locations suggest that the sail’s influence on pressure fluctuations diminishes with increasing distance from the sail due to the decay of large-scale turbulent eddies as they move away from the sail. Moreover, at points further downstream, such as $\mathrm{p}_5$ and $\mathrm{p}_6$, the influence of the appendage decreases. This is reflected in the PSD plots of $\mathrm{p}_5$ and $\mathrm{p}_6$ in Fig.~\ref{fig:suboff_surface_furessure}(b), where the frequency spectrum shows similar behavior, suggesting that the flow is transitioning towards a more turbulent state dominated by smaller-scale vortices.

Moreover, Fig.~\ref{fig:suboff_surface_furessure_fairwater} presents the time history and power spectral density of pressure fluctuations, providing critical insights into the flow dynamics around the SUBOFF model. The time history analysis, as depicted in Fig.~\ref{fig:suboff_surface_furessure_fairwater}(a, c, e), reveals significant variations in surface pressure fluctuations at different locations (points $\mathrm{p}_1$, $\mathrm{p}_2$, $\mathrm{p}_3$, and $\mathrm{p}_5$) near the fairwater and mid-hull. The standard deviation (SD) of the time history of the pressure fluctuations near the sail is markedly higher, with values of 347.39 Pa, 197.92 Pa, and 101.28 Pa for points $\mathrm{p}_1$, $\mathrm{p}_2$, and $\mathrm{p}_3$, respectively. This elevated SD is primarily driven by the complex turbulent vortex structures generated at the junction of the sail and hull, particularly the formation of horseshoe vortices, as discussed in Section~\ref{validation}. These vortices induce intense mixing and unsteady flow patterns, leading to pronounced pressure fluctuations in their vicinity. Specifically, the SD at $\mathrm{p}_1$, located closest to the sail-hull junction, is approximately 75\% higher than at $\mathrm{p}_2$ (347.39 Pa vs. 197.92 Pa), and at $\mathrm{p}_3$, it decreases by about 49\% compared to $\mathrm{p}_2$, reflecting the gradual attenuation of turbulence intensity with increasing distance from the sail-hull junction. In contrast, at point $\mathrm{p}_5$, located further downstream, the SD drops significantly to 49.69 Pa, representing an 85\% reduction compared to $\mathrm{p}_1$. This sharp decline underscores the diminishing influence of the sail-induced turbulence as the flow progresses downstream. The pronounced pressure fluctuations at points $\mathrm{p}_1$, $\mathrm{p}_2$, and $\mathrm{p}_3$ are directly linked to the formation and interaction of horseshoe vortices in the sail region, which creates localized regions of high-pressure fluctuation. In contrast, the smoother pressure signals at $\mathrm{p}_5$ indicate a more stabilized flow regime, where the effects of the sail's turbulence have largely dissipated. These findings align with the earlier discussions on turbulent vortex structures, emphasizing the critical role of the sail-hull junction in generating complex flow phenomena that significantly influence the surrounding pressure field.

Furthermore, from a frequency perspective, as shown in Fig.~\ref{fig:suboff_surface_furessure_fairwater}(b, d, f), the PSD of pressure fluctuations at points near the fairwater (e.g., $\mathrm{p}_1$) and along the top meridian line (e.g., $\mathrm{p}_5$) reveals a notable shift in the frequency spectrum. The differences between $\mathrm{p}_1$ (near the sail) and $\mathrm{p}_5$ (on the top surface of the hull in Fig.~\ref{fig:suboff_surface_furessure_fairwater}(b), far from the sail and in a fully developed turbulent boundary layer) are most pronounced in the mid-to-low frequency range ($\mathrm{St}<40$), where there is a difference of approximately 15-20 dB. This substantial difference is primarily linked to the large-scale turbulent and pressure structures generated by the interaction between the sail and hull, particularly the horseshoe vortex formed at the junction. The horseshoe vortex plays a pivotal role in the increased pressure fluctuations observed near the sail. In contrast, the differences between $\mathrm{p}_2$ (located near the sail but closer to $\mathrm{p}_5$) and $\mathrm{p}_5$ are smaller in Fig.~\ref{fig:suboff_surface_furessure_fairwater}(d), with a reduced difference of around 10 dB in the mid-to-low frequency range. However, a larger disparity of approximately 5 dB emerges in the high-frequency range, which is attributed to small-scale flow structures. This suggests that while the large-scale effects of the sail-hull interaction begin to attenuate as the flow moves downstream, the smaller turbulent structures persist, continuing to influence the pressure fluctuations at higher frequencies. Further downstream, at $\mathrm{p}_3$ (even closer to $\mathrm{p}_5$) in Fig.~\ref{fig:suboff_surface_furessure_fairwater}(f), the frequency spectrum shows greater similarity to $\mathrm{p}_5$, with only a minor upward shift of about 2 dB, indicating that the pressure fluctuations at $\mathrm{p}_3$ are largely dominated by the more developed turbulent boundary layer, with minimal influence from the sail-hull interaction. In summary, the significant differences observed in the frequency spectra, particularly in the mid-to-low frequency range, highlight the dominant role of the horseshoe vortex at the sail-hull junction in generating surface pressure fluctuations. As the flow moves downstream, the influence of this vortex structure diminishes, with the frequency spectrum gradually shifting towards the characteristics of a fully developed turbulent boundary layer.

\subsection{Far-field acoustic field}\label{Acoustic_field}
\begin{figure}[!htbp]
	\centering
	\begin{overpic}[trim=4cm 2cm 4cm 2cm, clip, width=6cm]{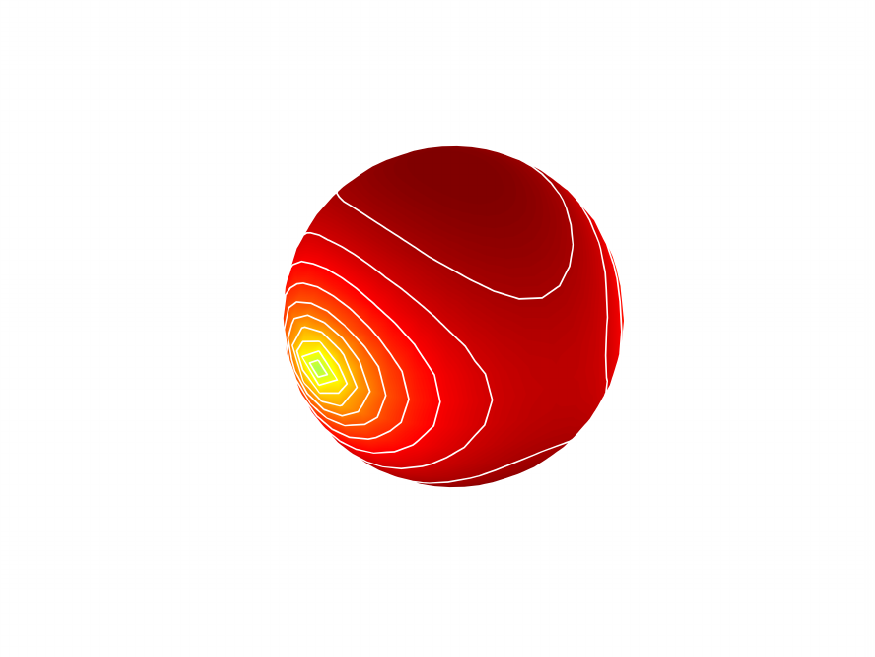}
    \put(-10,76){\color{black}{\textit{(a)}}}
    \put(70,10){\rotatebox{20}{\includegraphics[scale=0.05]{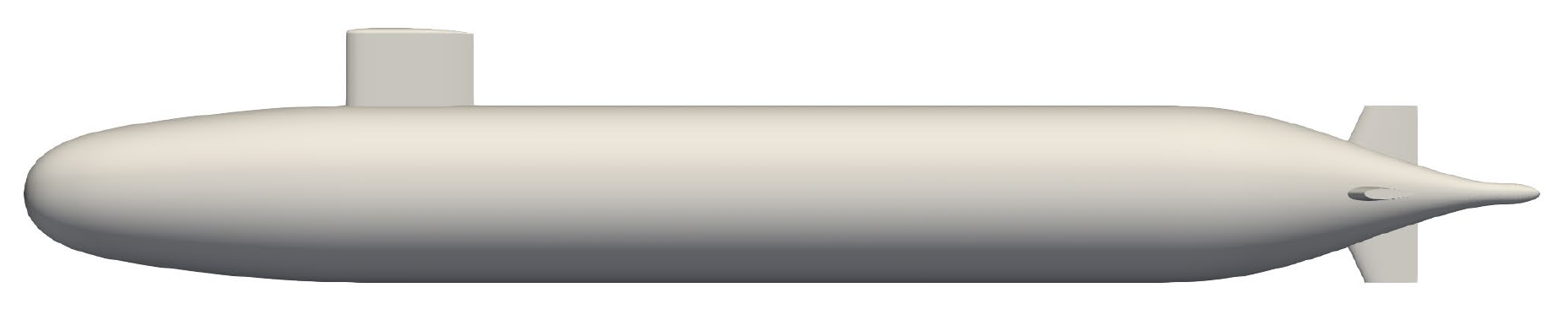}}}
	\end{overpic}
 \qquad  
	\begin{overpic}[trim=4cm 2cm 4cm 2cm, clip, width=6cm]{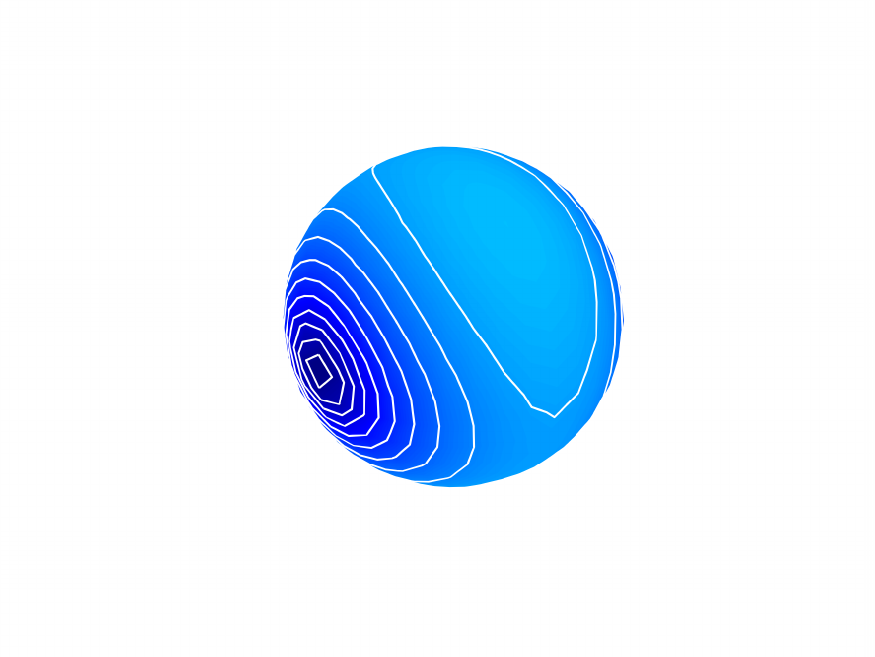}
    \put(-4,76){\color{black}{\textit{(b)}}}
    \put(70,10){\rotatebox{20}{\includegraphics[scale=0.05]{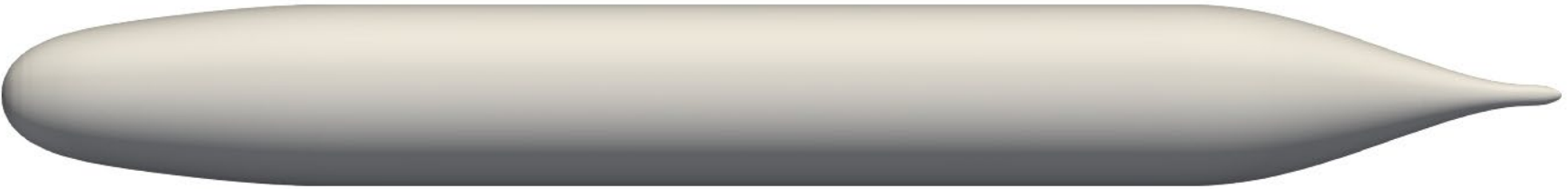}}}
    \put(130,40){\includegraphics[scale=0.5]{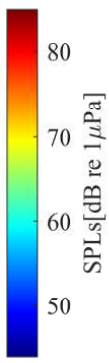}}
	\end{overpic}
	\caption[]{Comparison of the far-field \textit{SPL} across all spatial directions between (a) the appended SUBOFF and (b) the SUBOFF hull. Reference to the orientation of the SUBOFF model is given in the insets.}
	\label{fig:suboff_farfield_noise}
\end{figure}

\begin{figure}[!htbp]
	\centering
	\begin{overpic}[width=7cm]{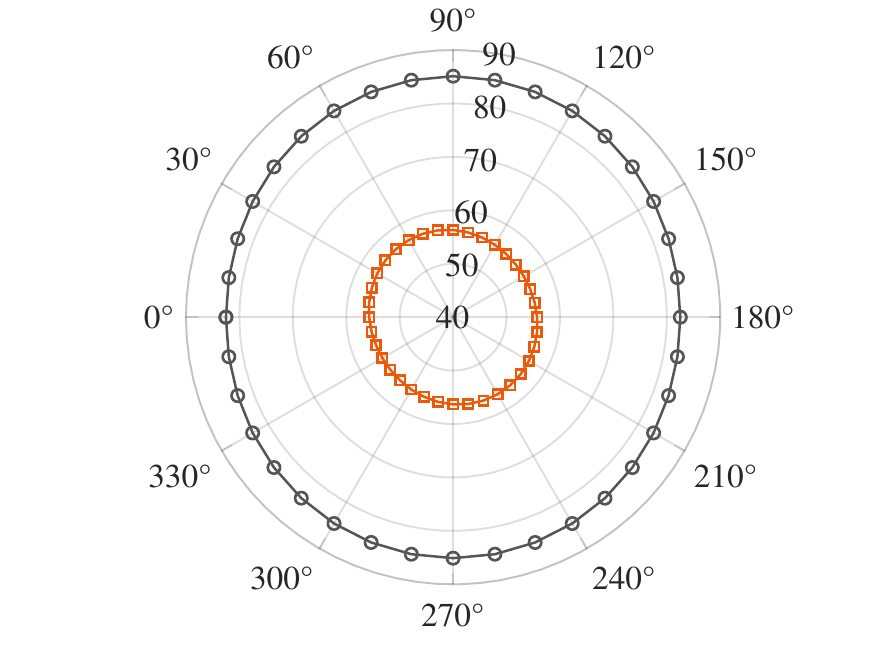}
    \put(0,70){\color{black}{\textit{(a)}}}
    \put(59,68){\rotatebox{-5}{\scriptsize\color{black}{[dB]}}}
	\end{overpic}
    \begin{overpic}[width=7cm]{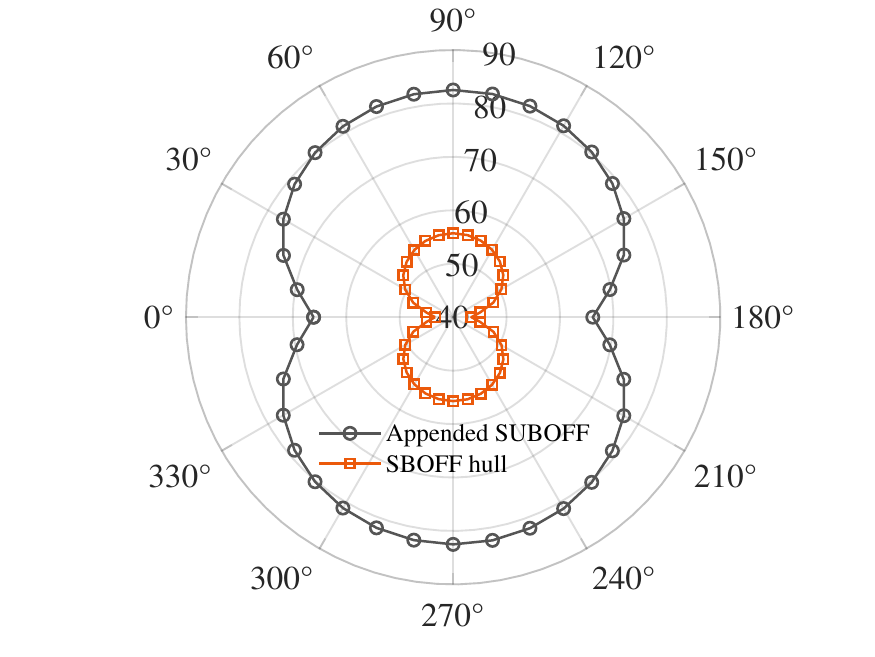}
    \put(0,70){\color{black}{\textit{(b)}}}
    \put(59,68){\rotatebox{-5}{\scriptsize\color{black}{[dB]}}}
	\end{overpic}
	\caption[]{Comparison of the far-field noise directivities between the appended SUBOFF model and the SUBOFF hull at the slice (a) $x=0$ and (b) $z=0$.}
	\label{fig:suboff_farfieldSPL}
\end{figure}

To derive the far-field spatial distribution of the flow noise, the sound field is calculated using 648 hydrophones positioned $500D$ from the model. This approach replicates the methodology employed in our previous study \cite{Jiang2024SUBOFF}. Fig.~\ref{fig:suboff_farfield_noise} presents three-dimensional views of the sound directivity for both the bare SUBOFF hull and the appended SUBOFF models. Significant differences are observed in the far-field \textit{SPLs} between the two models. The appended SUBOFF exhibits far-field \textit{SPLs} approximately 20 dB higher than those of the bare SUBOFF hull, highlighting the substantial influence of the appendages on the model’s acoustic characteristics. For the appended SUBOFF, the sound pressure reaches its minimum value of 66.11 dB at the locations $(\pm500D,0,0)$, while it peaks at 85.10 dB on the $yoz$ plane. In contrast, the bare SUBOFF hull demonstrates a more uniform distribution of acoustic pressure across these directions, with its \textit{SPL} remaining relatively constant at around 50-55 dB in most directions. These results further suggest that the appendages cause more directional sound radiation, especially noticeable in the upstream and downstream directions in the $x$ axis.

Additionally, Fig.~\ref{fig:suboff_farfieldSPL} presents a polar plot of the far-field sound pressure level. The directivity plot for the vertical plane at $z/D=0$ reveals a stronger dipole pattern in the appended SUBOFF compared to the bare hull. This pronounced dipole effect underscores the significant impact of the appendages on acoustic emission, especially in terms of enhancing pressure fluctuations at specific points. While both models exhibit some general similarities in their acoustic pressure distributions, such as comparable levels on the port and starboard sides, as well as in the upstream and downstream directions, these similarities are overshadowed by the notable differences in \textit{SPLs} and directivity patterns between the two models.

\begin{figure*}[!htbp]
	\centering
	\begin{overpic}[width=6cm]{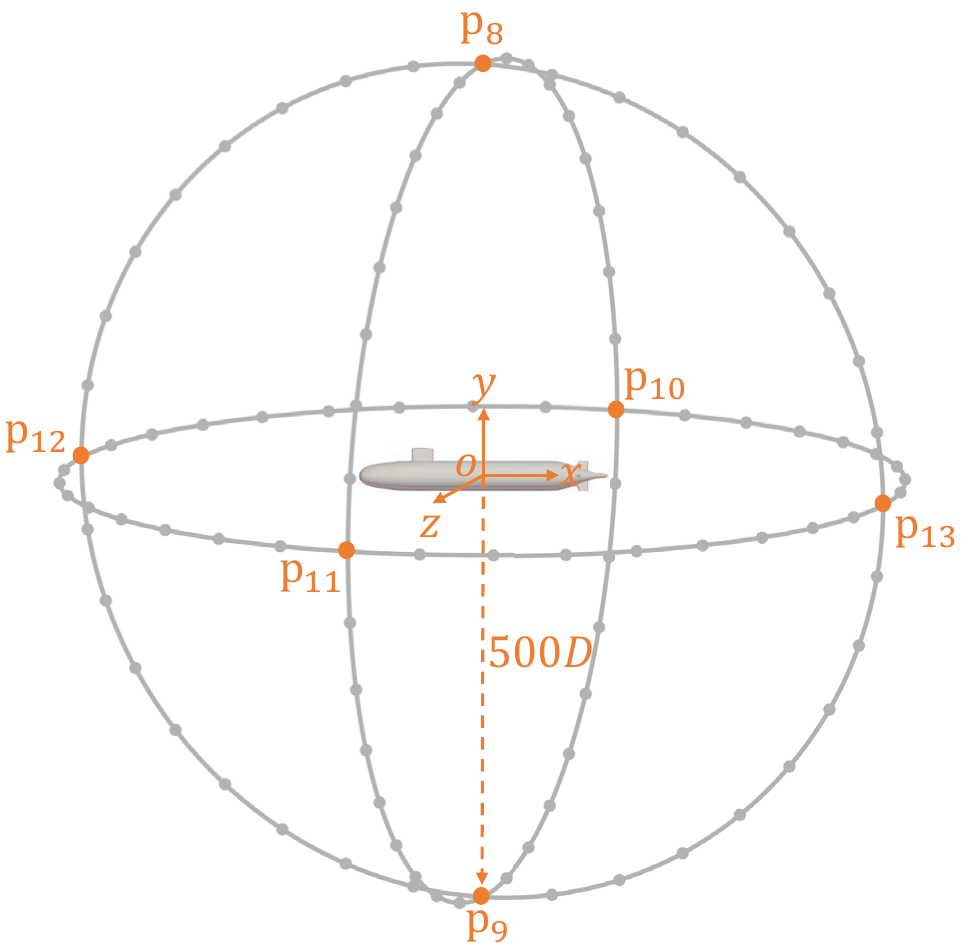}
        \put(-5,85){\color{black}{\textit{(a)}}}
	\end{overpic}
 	\begin{overpic}[width=14cm]{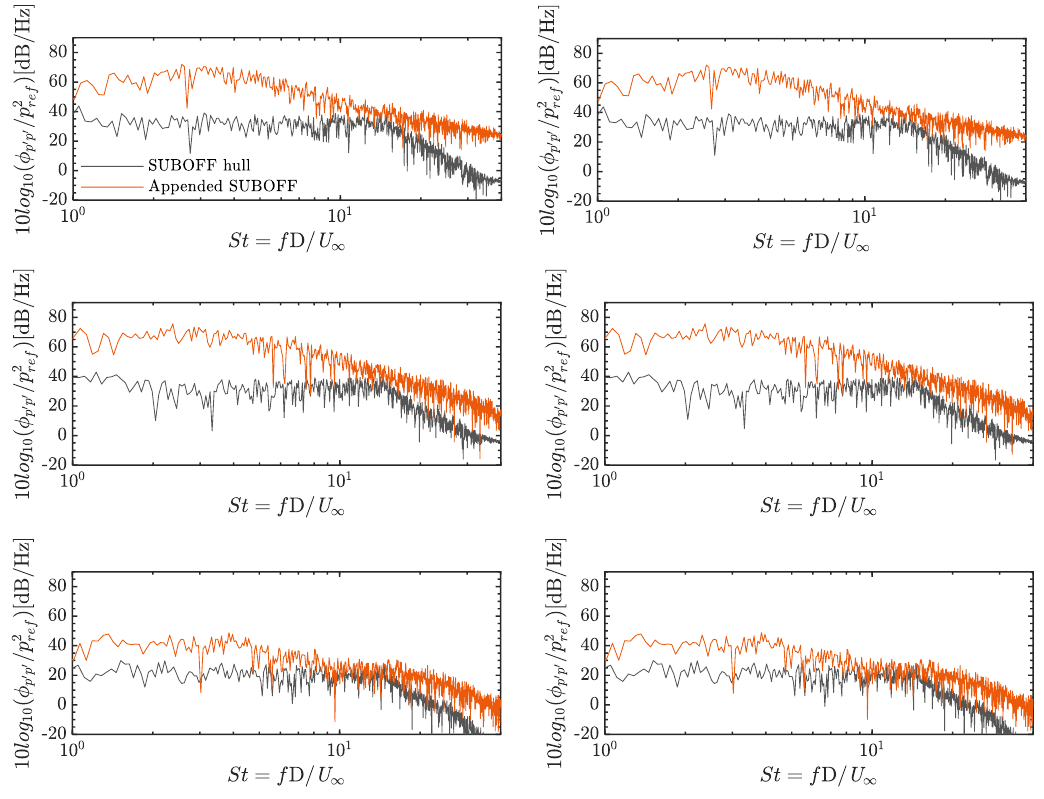}
        \put(44,69){{$\mathrm{p}_8$}}
        \put(94,69){{$\mathrm{p}_9$}}
        \put(44,45){{$\mathrm{p}_{10}$}}
        \put(94,45){{$\mathrm{p}_{11}$}}
        \put(44,19){{$\mathrm{p}_{12}$}}
        \put(94,19){{$\mathrm{p}_{13}$}}
        \put(-3,73){\color{black}{\textit{(b)}}}
        \put(48,73){\color{black}{\textit{(c)}}}
        
        \put(-3,48){\color{black}{\textit{(d)}}}
        \put(48,48){\color{black}{\textit{(e)}}}

        \put(-3,22){\color{black}{\textit{(f)}}}
        \put(48,22){\color{black}{\textit{(g)}}}
	\end{overpic}
	\caption[]{\textcolor{black}{(a) Far-field noise observers of the SUBOFF model, (b-g) the PSD of the far-field sound pressure
fluctuations at different locations on the appended SUBOFF model compared with that on the SUBOFF hull from $\mathrm{p_{8}}$ to $\mathrm{p_{13}}$}.}
	\label{fig:observers}
\end{figure*}
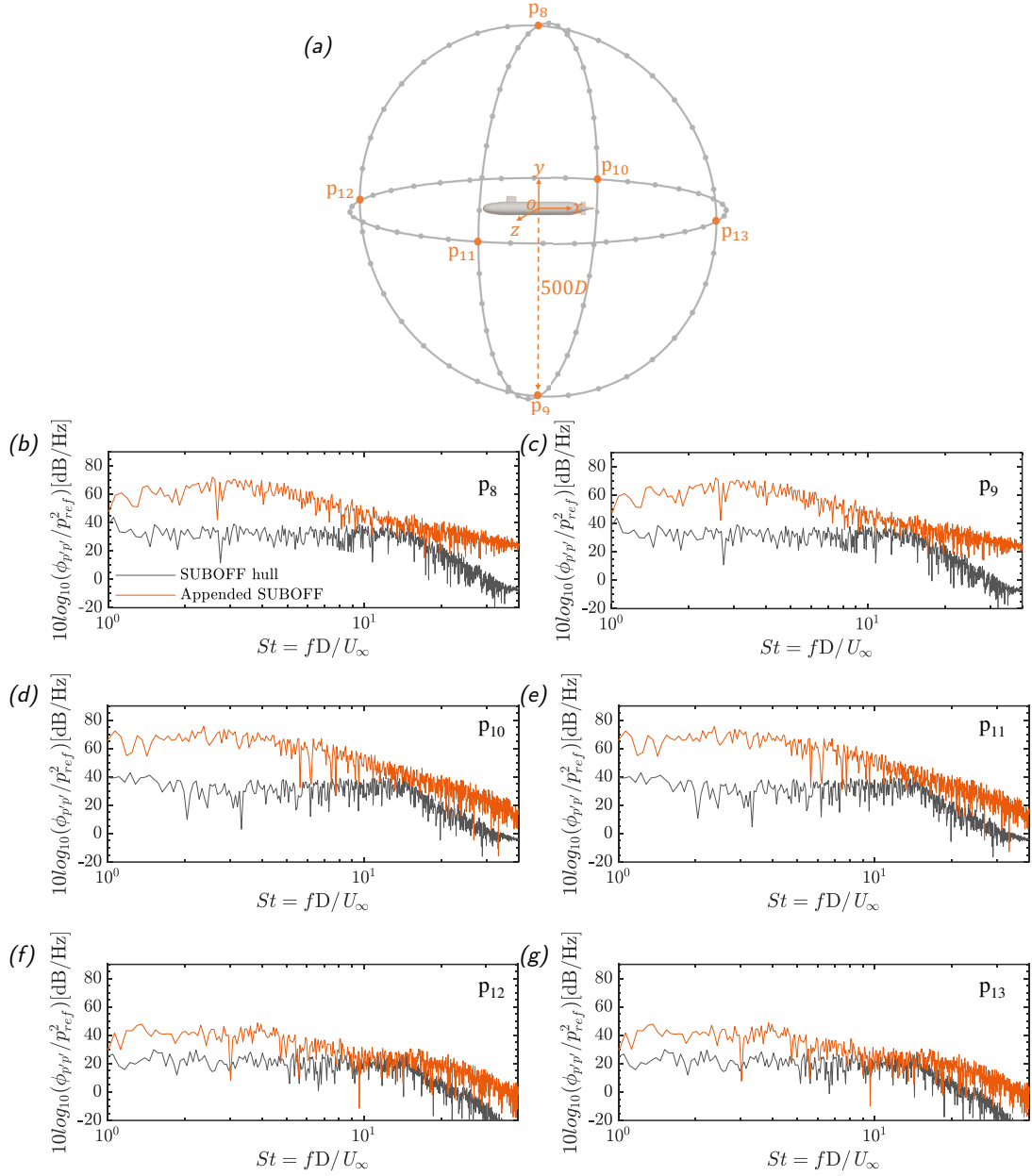

As observed in Fig.~\ref{fig:suboff_farfield_noise}, the appended SUBOFF consistently exhibits higher sound pressure levels across all spatial directions compared to the SUBOFF hull. This difference can be attributed to the increased complexity of turbulent structures introduced by the appendages. These structures lead to increased sound emission, particularly at specific frequency ranges, which can be better understood by analyzing the frequency content of the PSD curves. Thus, we illustrate the distribution of sound pressure levels in the frequency domain, with corresponding PSD curves between two SUBOFF configurations: the appended SUBOFF and the SUBOFF hull. These measurements are taken at several far-field locations: $\mathrm{p_8}(0,500D,0)$, $\mathrm{p_9}(0,-500D,0)$, $\mathrm{p_{10}}(0,0,-500D)$, $\mathrm{p_{11}}(0,0,500D)$, $\mathrm{p_{12}}(-500D,0,0)$, and $\mathrm{p_{13}}(500D,0,0)$. As shown in Fig.~\ref{fig:observers}, the appended SUBOFF model produces a broader range of acoustic emissions across both low and high frequencies compared to the bare hull. For instance, at low frequencies (St < 10), the appended SUBOFF model exhibits higher sound levels (about 15-20 dB increase) than the bare hull. At higher frequencies, particularly in the range between 10 < St < 50,  the frequency spectrum for the appended SUBOFF shows a decrease in sound pressure levels relative to the low-frequency range but remains higher than that of the bare hull. 

The relationship between the observed differences in PSD plots and flow dynamics can be interpreted in terms of the scale of the sound sources: 
(i) For the low-frequency content increase of the appended SUBOFF model: The appended SUBOFF exhibits higher sound levels at low frequencies (St < 10), indicating the dominance of large-scale turbulent structures. These structures are likely associated with large-scale vortex generation and shedding from the appendages, particularly the sail and fins, such as the horseshoe vortex and tip vortices, which generate low-frequency pressure fluctuations detectable in the far-field. The presence of these large vortices is directly linked to the strong pressure fluctuations observed at low frequencies.
(ii) For the high-frequency content increase of the appended SUBOFF model: As the frequency increases (St > 10), the sound pressure levels at the appended SUBOFF remain higher than those of the bare hull. This behavior can be attributed to the enhanced small-scale turbulent eddies that form in the wake of the appendages. These small-scale vortices, with higher wavenumbers, are responsible for generating higher-frequency noise. The contribution from these eddies is significant at points located downstream of the appendages, mainly at the middle axisymmetrical body, where the influence of the appendages is more pronounced at higher frequencies.

It is important to note that the frequency spectrum of far-field noise in Fig.~\ref{fig:suboff_farfield_noise} does not directly match the frequency spectrum of pressure fluctuations at individual surface measurement points in Figs.~\ref{fig:suboff_surface_furessure} and \ref{fig:suboff_surface_furessure_fairwater}. This difference is reasonable due to the fundamental differences in how these spectra are derived. The surface pressure fluctuations measured at specific points on the submarine's surface represent the pressure disturbances at those locations, which are closely related to the local turbulent structures and vortex shedding occurring at the wall. These measurements are localized and reflect the pressure field at specific points where the turbulent structures have a direct contribution to the acoustic emission. In contrast, the far-field noise is calculated using the FW-H equation, which integrates the pressure fluctuations over the entire submarine surface. This method accounts for all the sound sources on the surface, including those generated by the appendages, and integrates their contributions to calculate the overall far-field noise. Thus, the far-field acoustic spectrum is the result of combining the effects of all turbulent structures (both large-scale and small-scale) across the entire surface of the submarine. The FW-H formulation provides a more holistic representation of the sound field, incorporating the interactions between the various turbulent sources across the model's surface.

\begin{table*}[width=0.8\linewidth,cols=7,pos=!htbp]
\caption{\textcolor{black}{Comparison of the overall sound pressure level at different far-field observers.}}
\label{tab:suboff_OASPL}
\begin{tabular*}{\tblwidth}{@{} CCCCCCC@{} }
\toprule
\multirow{2}{*}{Model} & \multicolumn{6}{c}{OASPL [dB] at different Observers} \\
& {$\mathrm{p_{8}}$} & {$\mathrm{p_{9}}$} & {$\mathrm{p_{10}}$} & {$\mathrm{p_{11}}$} & {$\mathrm{p_{12}}$} & {$\mathrm{p_{13}}$} \\ \midrule
Appended SUBOFF          & 94.09 & 94.09 & 93.83 & 93.83 & 82.65 & 82.71 \\
SUBOFF hull              & 82.54 & 82.53 & 85.12 & 85.11 & 66.13 & 66.15 \\ 
\bottomrule
\end{tabular*}
\end{table*}

Moreover, the overall sound pressure levels (OASPL) at these far-field observers are summarized in Table~\ref{tab:suboff_OASPL}. OASPL is a metric used to quantify the total acoustic energy of a sound signal across all frequencies. It is expressed in dB and provides a single value representing the cumulative sound pressure level over the entire frequency spectrum. The OASPL difference between the appended SUBOFF and SUBOFF hull varies across observers, with an average difference of approximately 10 dB. This average difference reflects a significant impact on the acoustic field due to the appendages, aligning with the theory that more complex flow patterns (e.g., separated flows, vortex shedding) increase the noise levels generated by the body. For instance, at $\mathrm{p_8}$, the \textit{SPL} for the appended SUBOFF is 94.09 dB, whereas the SUBOFF hull's \textit{SPL} is only 82.54 dB. This 11.55 dB difference confirms that the added appendages significantly affect the model's sound radiation. These results confirm that the appendages significantly alter the flow structure, enhancing both the large-scale and small-scale turbulent noise generation.

\section{Conclusions}\label{conclusions}
This study investigates the influence of appendages on the turbulent flow and acoustic characteristics of the DARPA SUBOFF model at a Reynolds number of \(1.2 \times 10^7\). High-fidelity simulations are conducted using
WMLES and FW-H acoustic analogy, implemented with a high-order numerical scheme. The model is first validated against available experimental and numerical data, demonstrating excellent agreement in predicting pressure and drag coefficients, thereby confirming its capability to accurately capture the hydrodynamic forces acting on the SUBOFF model. The effects of appendages are examined through three key aspects.
First, the hydrodynamic forces and turbulent flow characteristics are analyzed, including the components of hydrodynamic forces, mean surface pressure distribution, near-wall velocity and pressure profiles in the stern region, and turbulent coherent structures for both the appended SUBOFF and the axisymmetric hull. Second, surface pressure fluctuations and noise generation are evaluated through the time history and spectral analysis of pressure fluctuations at multiple locations on the appended SUBOFF and its bare hull. Third, the far-field acoustic pressure distribution is studied by comparing sound pressure levels between the two SUBOFF configurations and assessing the overall sound pressure levels at far-field observer positions. Additionally, 648 hydrophones positioned at $500D$ from the model are employed to investigate the far-field sound distribution. The main findings of this study are as follows:
\begin{enumerate}[(i).]
\item \textbf{Hydrodynamic Effects of Appendages:} The drag force on the appended SUBOFF model is dominated by viscous forces, similar to the bare hull. However, appendages significantly amplify pressure-induced contributions due to localized flow separations and vortex shedding. The sail significantly impacts the mean pressure and velocity over the mid-body, while the fins disrupt the velocity and pressure profiles in their wake, affecting the outer region of the near-wall flow.
Both geometries share features like turbulent boundary layer development and stern flow separation, but appendages introduce complex vortical structures (e.g., horseshoe, hairpin, and necklace vortices) at junctions and downstream regions.

\item \textbf{Surface Pressure Fluctuations:} The presence of appendages leads to a marked increase in surface pressure fluctuations, with the sail-hull junction acting as a dominant source of these fluctuations. The PSDs of surface pressure fluctuations reveal an order-of-magnitude increase in amplitude near the sail, particularly in the low-frequency range, corresponding to large-scale turbulent structures. The sail’s influence is most pronounced in the upper part of the hull and diminishes downstream. Notably, the sail's effect does not significantly extend to the lower hull, confirming the spatial variability of appendage-induced disturbances. The frequency analysis further emphasizes the decreasing influence of the sail on pressure fluctuations as the distance from the sail increases, highlighting the complex flow dynamics influenced by the appendages.

\item \textbf{Far-field Acoustic Characteristics:} The far-field \textit{SPLs} for the appended SUBOFF model are approximately 20 dB higher than for the bare hull, with a pronounced dipole directivity pattern. The sound pressure peaks at 85.10 dB on the central plane, while minima (66.11 dB) occurs along the streamwise direction. The OASPLs for the appended model are higher by approximately 10 dB compared to the bare hull, underscoring the significant acoustic impact of the appendages. The far-field acoustic spectra show broader fluctuations and directional asymmetry due to the complex turbulent structures induced by the appendages. Both configurations exhibit similar spectral trends, but the appended SUBOFF model introduces higher sound levels at both low and high frequencies, reflecting the combined effects of large-scale vortex dynamics and small-scale turbulent eddies.
\end{enumerate}

In conclusion, this study provides comprehensive data on the far-field noise distribution for the appended high Reynolds number SUBOFF model, filling a significant gap in the current literature regarding far-field turbulence noise for high Reynolds number submarine configurations. The results highlight the crucial role of appendages in shaping both the hydrodynamic and acoustic behavior of the model. These findings offer new insights into turbulence noise generation for complex underwater geometries and lay the groundwork for future submarine design efforts that aim to optimize hydrodynamic and acoustic performance, ultimately enhancing environmental and engineering outcomes.

\section*{Acknowledgements}
This work was supported in part by National Natural Science Foundation of China (Grant Nos. 12472293, 91752104), the Shanghai Rising-Star Program (No. 23QA1405000), the Fundamental Research Funds for the Central Universities (No. AF0100142) as well as the fund from Shanghai Pilot Program for Basic Research - Shanghai Jiao Tong University (No. 21TQ1400202).

\printcredits

\section*{Data availability statement}
The data that support the findings of this study are available from the corresponding author upon reasonable request.

\section*{Author ORCIDs}
\noindent Peng Jiang, \href{https://orcid.org/0000-0002-9072-8307}{https://orcid.org/0000-0002-9072-8307};\\
Shijun Liao, \href{https://orcid.org/0000-0002-2372-9502}{https://orcid.org/0000-0002-2372-9502};\\
Ling Liu, \href{https://orcid.org/0000-0002-5146-8063}{https://orcid.org/0000-0002-5146-8063};\\
Bin Xie, \href{https://orcid.org/0000-0002-4218-2442}{https://orcid.org/0000-0002-4218-2442}.


\section*{Declaration of competing interest}
The authors report no conflict of interest. The authors declare that they have no known competing financial interests or personal relationships that could have appeared to influence the work reported in this paper.

\bibliographystyle{cas-model2-names}

\begin{thebibliography}{48}
\expandafter\ifx\csname natexlab\endcsname\relax\def\natexlab#1{#1}\fi
\providecommand{\url}[1]{\texttt{#1}}
\providecommand{\href}[2]{#2}
\providecommand{\path}[1]{#1}
\providecommand{\DOIprefix}{doi:}
\providecommand{\ArXivprefix}{arXiv:}
\providecommand{\URLprefix}{URL: }
\providecommand{\Pubmedprefix}{pmid:}
\providecommand{\doi}[1]{\href{http://dx.doi.org/#1}{\path{#1}}}
\providecommand{\Pubmed}[1]{\href{pmid:#1}{\path{#1}}}
\providecommand{\bibinfo}[2]{#2}
\ifx\xfnm\relax \def\xfnm[#1]{\unskip,\space#1}\fi
\bibitem[{Chen et~al.(2023)Chen, Yang, Zhao and Wan}]{Chen2023SUBOFF}
\bibinfo{author}{Chen, S.T.}, \bibinfo{author}{Yang, L.C.}, \bibinfo{author}{Zhao, W.W.}, \bibinfo{author}{Wan, D.C.}, \bibinfo{year}{2023}.
\newblock \bibinfo{title}{\text{Wall-modeled} large eddy simulation for the flows around an axisymmetric body of revolution}.
\newblock \bibinfo{journal}{J.~Hydrodyn.} \bibinfo{volume}{35}, \bibinfo{pages}{199--209}.
\bibitem[{Choi and Moin(2012)}]{Choi2012gridWMLES}
\bibinfo{author}{Choi, H.}, \bibinfo{author}{Moin, P.}, \bibinfo{year}{2012}.
\newblock \bibinfo{title}{\text{Grid-point} requirements for large eddy simulation: {Chapman's} estimates revisited}.
\newblock \bibinfo{journal}{Phys.~Fluids} \bibinfo{volume}{24}, \bibinfo{pages}{011702}.
\bibitem[{Chorin(1968)}]{Chorin1968Numerical}
\bibinfo{author}{Chorin, A.J.}, \bibinfo{year}{1968}.
\newblock \bibinfo{title}{\text{Numerical} solution of the {Navier-Stokes} equations}.
\newblock \bibinfo{journal}{Math.~Comput.} \bibinfo{volume}{22}, \bibinfo{pages}{745--762}.
\bibitem[{Cianferra and Armenio(2021)}]{Cianferra2021}
\bibinfo{author}{Cianferra, M.}, \bibinfo{author}{Armenio, V.}, \bibinfo{year}{2021}.
\newblock \bibinfo{title}{\text{Scaling} properties of the {Ffowcs-Williams and Hawkings} equation for complex acoustic source close to a free surface}.
\newblock \bibinfo{journal}{J. Fluid Mech.} \bibinfo{volume}{927}, \bibinfo{pages}{A2}.
\bibitem[{Crook(1990)}]{Crook1990}
\bibinfo{author}{Crook, L.B.}, \bibinfo{year}{1990}.
\newblock \bibinfo{title}{\text{Resistance} for {DARPA SUBOFF} as Represented by Model 5470}.
\newblock \bibinfo{type}{Ship Hydrodynamics Report} \bibinfo{number}{DTRC/SHD-1298-07}. David Taylor Research Center.
\bibitem[{Duprat et~al.(2011)Duprat, Balarac, Métais, Congedo and Brugière}]{Duprat2011nutmodel}
\bibinfo{author}{Duprat, C.}, \bibinfo{author}{Balarac, G.}, \bibinfo{author}{Métais, O.}, \bibinfo{author}{Congedo, P.}, \bibinfo{author}{Brugière, O.}, \bibinfo{year}{2011}.
\newblock \bibinfo{title}{\text{A} wall-layer model for large-eddy simulations of turbulent flows with/out pressure gradient}.
\newblock \bibinfo{journal}{Phys.~Fluids} \bibinfo{volume}{23}, \bibinfo{pages}{015101}.
\bibitem[{{Ffowcs Williams} and Hawkings(1969)}]{FWH1969}
\bibinfo{author}{{Ffowcs Williams}, J.E.}, \bibinfo{author}{Hawkings, D.L.}, \bibinfo{year}{1969}.
\newblock \bibinfo{title}{\text{Sound} generation by turbulence and surfaces in arbitrary motion}.
\newblock \bibinfo{journal}{Philos.~Trans.~R.~Soc.~London~Ser.~A} \bibinfo{volume}{264}, \bibinfo{pages}{321--342}.
\bibitem[{Gottlieb and Shu(1998)}]{Gottlieb1998TVD}
\bibinfo{author}{Gottlieb, S.}, \bibinfo{author}{Shu, C.W.}, \bibinfo{year}{1998}.
\newblock \bibinfo{title}{\text{Total} variation diminishing {Runge-Kutta} schemes}.
\newblock \bibinfo{journal}{Math.~Comput.} \bibinfo{volume}{67}, \bibinfo{pages}{73--85}.
\bibitem[{Groves et~al.(1989)Groves, Huang and Chang}]{groves1989geometric}
\bibinfo{author}{Groves, N.C.}, \bibinfo{author}{Huang, T.T.}, \bibinfo{author}{Chang, M.S.}, \bibinfo{year}{1989}.
\newblock \bibinfo{title}{{Geometric characteristics of DARPA (Defense Advanced Research Projects Agency) SUBOFF models (DTRC model numbers 5470 and 5471)}}.
\newblock \bibinfo{type}{SHD-1298-01}. David Taylor Research Center.
\bibitem[{He et~al.(2024)He, Pan, Zhao, Wang and Wan}]{He2024Review}
\bibinfo{author}{He, K.J.}, \bibinfo{author}{Pan, Z.}, \bibinfo{author}{Zhao, W.W.}, \bibinfo{author}{Wang, J.H.}, \bibinfo{author}{Wan, D.C.}, \bibinfo{year}{2024}.
\newblock \bibinfo{title}{\text{Overview} of research progress on numerical simulation methods for turbulent flows around underwater vehicles}.
\newblock \bibinfo{journal}{J.~Marine.~Sci.~Appl.} \bibinfo{volume}{23}, \bibinfo{pages}{1--22}.
\bibitem[{Holloway et~al.(2015)Holloway, Jeans and Watt}]{Holloway2015}
\bibinfo{author}{Holloway, A.G.L.}, \bibinfo{author}{Jeans, T.L.}, \bibinfo{author}{Watt, G.D.}, \bibinfo{year}{2015}.
\newblock \bibinfo{title}{\text{Flow} separation from submarine shaped bodies of revolution in steady turning}.
\newblock \bibinfo{journal}{Ocean Eng.} \bibinfo{volume}{108}, \bibinfo{pages}{426--438}.
\bibitem[{Hu et~al.(2023)Hu, Hayat and Park}]{Hu2023WMLES}
\bibinfo{author}{Hu, X.}, \bibinfo{author}{Hayat, I.}, \bibinfo{author}{Park, G.I.}, \bibinfo{year}{2023}.
\newblock \bibinfo{title}{\text{Wall-modelled} large-eddy simulation of three-dimensional turbulent boundary layer in a bent square duct}.
\newblock \bibinfo{journal}{J.~Fluid Mech.} \bibinfo{volume}{960}, \bibinfo{pages}{A29}.
\bibitem[{Huang et~al.(1992)Huang, Liu, Groves, Forlini, Blanton and Gowing}]{Huang1992Exp}
\bibinfo{author}{Huang, T.}, \bibinfo{author}{Liu, H.L.}, \bibinfo{author}{Groves, N.C.}, \bibinfo{author}{Forlini, T.}, \bibinfo{author}{Blanton, J.}, \bibinfo{author}{Gowing, S.}, \bibinfo{year}{1992}.
\newblock \bibinfo{title}{\text{Measurements} of flows over an axisymmetric body with various appendages in a wind tunnel: {The DARPA SUBOFF} experimental program}, in: \bibinfo{booktitle}{In Proceedings of the 19th Symposium on Naval Hydrodynamics}.
\bibitem[{Jiang et~al.(2025)Jiang, Huang, Cao, Liao and Xie}]{Jiang2024Sphere}
\bibinfo{author}{Jiang, P.}, \bibinfo{author}{Huang, Y.C.}, \bibinfo{author}{Cao, Y.}, \bibinfo{author}{Liao, S.J.}, \bibinfo{author}{Xie, B.}, \bibinfo{year}{2025}.
\newblock \bibinfo{title}{\text{Utility} of high-order scheme for unsteady flow simulations: Comparison with second-order tool}.
\newblock \bibinfo{journal}{J. Ocean. Eng. Sci.} .
\bibitem[{Jiang et~al.(2024)Jiang, Liao and Xie}]{Jiang2024SUBOFF}
\bibinfo{author}{Jiang, P.}, \bibinfo{author}{Liao, S.J.}, \bibinfo{author}{Xie, B.}, \bibinfo{year}{2024}.
\newblock \bibinfo{title}{\text{Large-eddy} simulation of flow noise from turbulent flows past an axisymmetric hull using high-order schemes}.
\newblock \bibinfo{journal}{Ocean Eng.} \bibinfo{volume}{312}, \bibinfo{pages}{119150}.
\bibitem[{Jim{\'e}nez et~al.(2010a)Jim{\'e}nez, Hultmark and Smits}]{Jimenez2010}
\bibinfo{author}{Jim{\'e}nez, J.M.}, \bibinfo{author}{Hultmark, M.}, \bibinfo{author}{Smits, A.J.}, \bibinfo{year}{2010}a.
\newblock \bibinfo{title}{\text{The} intermediate wake of a body of revolution at high {Reynolds} numbers}.
\newblock \bibinfo{journal}{J.~Fluid Mech.} \bibinfo{volume}{659}, \bibinfo{pages}{516--539}.
\bibitem[{Jim{\'e}nez et~al.(2010b)Jim{\'e}nez, Reynolds and Smits}]{Jimenez2010Effects}
\bibinfo{author}{Jim{\'e}nez, J.M.}, \bibinfo{author}{Reynolds, R.T.}, \bibinfo{author}{Smits, A.J.}, \bibinfo{year}{2010}b.
\newblock \bibinfo{title}{\text{The} effects of fins on the intermediate wake of a submarine model}.
\newblock \bibinfo{journal}{J.~Fluids~Eng.} \bibinfo{volume}{132}.
\bibitem[{Kim and Moin(1985)}]{Kim1985Application}
\bibinfo{author}{Kim, J.}, \bibinfo{author}{Moin, P.}, \bibinfo{year}{1985}.
\newblock \bibinfo{title}{\text{Application} of a fractional-step method to incompressible {Navier-Stokes} equations}.
\newblock \bibinfo{journal}{J.~Comput.~Phys.} \bibinfo{volume}{59}, \bibinfo{pages}{308--323}.
\bibitem[{Kumar and Mahesh(2018)}]{Kumar2018SUBOFF}
\bibinfo{author}{Kumar, P.}, \bibinfo{author}{Mahesh, K.}, \bibinfo{year}{2018}.
\newblock \bibinfo{title}{\text{Large-eddy} simulation of flow over an axisymmetric body of revolution}.
\newblock \bibinfo{journal}{J.~Fluid Mech.} \bibinfo{volume}{853}, \bibinfo{pages}{537--563}.
\bibitem[{Leong et~al.(2016)Leong, Piccolin, Desjuzeur et~al.}]{Leong2016}
\bibinfo{author}{Leong, Z.Q.}, \bibinfo{author}{Piccolin, S.}, \bibinfo{author}{Desjuzeur, M.}, et~al., \bibinfo{year}{2016}.
\newblock \bibinfo{title}{\text{Evaluation} of the out-of-plane loads on a submarine undergoing a steady turn}, in: \bibinfo{booktitle}{20th Australasian Fluid Mechanics Conference}, pp. \bibinfo{pages}{1--4}.
\bibitem[{Li et~al.(2016)Li, Liu, Wu and Chen}]{Li2016review}
\bibinfo{author}{Li, H.}, \bibinfo{author}{Liu, C.W.}, \bibinfo{author}{Wu, F.l.}, \bibinfo{author}{Chen, C.}, \bibinfo{year}{2016}.
\newblock \bibinfo{title}{\text{A} review of the progress for computational methods of hydrodynamic noise}.
\newblock \bibinfo{journal}{J.~Ship Res.} \bibinfo{volume}{11}, \bibinfo{pages}{72--89}.
\newblock \bibinfo{note}{(in Chinese)}.
\bibitem[{Liu et~al.(2023a)Liu, Hao, Bie, Wang, Ren and Hua}]{Liu2023suboffVortex}
\bibinfo{author}{Liu, G.}, \bibinfo{author}{Hao, Z.}, \bibinfo{author}{Bie, H.}, \bibinfo{author}{Wang, Y.}, \bibinfo{author}{Ren, W.}, \bibinfo{author}{Hua, Z.}, \bibinfo{year}{2023}a.
\newblock \bibinfo{title}{\text{Control} mechanism of a vortex control baffle for the horseshoe vortex around the sail of a {DARPA SUBOFF} model}.
\newblock \bibinfo{journal}{Ocean Eng.} \bibinfo{volume}{275}, \bibinfo{pages}{114166}.
\bibitem[{Liu and Huang(1998)}]{Liu1998SUBOFFexp}
\bibinfo{author}{Liu, H.}, \bibinfo{author}{Huang, T.}, \bibinfo{year}{1998}.
\newblock \bibinfo{title}{\text{Summary} of DARPA SUBOFF Experimental Program Data}.
\newblock \bibinfo{type}{Report} \bibinfo{number}{CRDKNSWC/HD-1298-11}. Naval Surface Warfare Center, Carderock Division, Hydrodynamics Directorate. \bibinfo{address}{Bethesda, MD}.
\bibitem[{Liu et~al.(2021)Liu, Chen, Yu, Zhang and Wang}]{Liu2021URANS}
\bibinfo{author}{Liu, L.W.}, \bibinfo{author}{Chen, M.X.}, \bibinfo{author}{Yu, J.W.}, \bibinfo{author}{Zhang, Z.G.}, \bibinfo{author}{Wang, X.Z.}, \bibinfo{year}{2021}.
\newblock \bibinfo{title}{\text{Full-scale} simulation of self-propulsion for a free-running submarine}.
\newblock \bibinfo{journal}{Phys.~Fluids} \bibinfo{volume}{33}, \bibinfo{pages}{047103}.
\bibitem[{Liu et~al.(2023b)Liu, Wang, Wang and He}]{LIU2023112009}
\bibinfo{author}{Liu, Y.}, \bibinfo{author}{Wang, H.P.}, \bibinfo{author}{Wang, S.Z.}, \bibinfo{author}{He, G.W.}, \bibinfo{year}{2023}b.
\newblock \bibinfo{title}{\text{A} cache-efficient reordering method for unstructured meshes with applications to wall-resolved large-eddy simulations}.
\newblock \bibinfo{journal}{J.~Comput.~Phys.} \bibinfo{volume}{480}, \bibinfo{pages}{112009}.
\bibitem[{Liu et~al.(2011)Liu, Xiong and Tu}]{Liu2011}
\bibinfo{author}{Liu, Z.H.}, \bibinfo{author}{Xiong, Y.}, \bibinfo{author}{Tu, C.X.}, \bibinfo{year}{2011}.
\newblock \bibinfo{title}{\text{Numerical} simulation and control of horseshoe vortex around an appendage-body junction}.
\newblock \bibinfo{journal}{J. Fluids Struct.} \bibinfo{volume}{27}, \bibinfo{pages}{23--42}.
\bibitem[{Ma et~al.(2025)Ma, Li, Guo, Liao and Yang}]{Ma2025JFS}
\bibinfo{author}{Ma, Z.H.}, \bibinfo{author}{Li, P.}, \bibinfo{author}{Guo, H.}, \bibinfo{author}{Liao, K.}, \bibinfo{author}{Yang, Y.R.}, \bibinfo{year}{2025}.
\newblock \bibinfo{title}{Performance and mechanism of the hydrodynamic noise reduction for biomimetic trailing-edge serrations of a submarine}.
\newblock \bibinfo{journal}{J. Fluids Struct.} \bibinfo{volume}{133}, \bibinfo{pages}{104256}.
\bibitem[{Ma et~al.(2024)Ma, Li, Wang, Lu and Yang}]{Ma2024OE}
\bibinfo{author}{Ma, Z.H.}, \bibinfo{author}{Li, P.}, \bibinfo{author}{Wang, L.Z.}, \bibinfo{author}{Lu, J.}, \bibinfo{author}{Yang, Y.R.}, \bibinfo{year}{2024}.
\newblock \bibinfo{title}{Mechanistic study of noise source and propagation characteristics of flow noise of a submarine}.
\newblock \bibinfo{journal}{Ocean Eng.} \bibinfo{volume}{302}, \bibinfo{pages}{117667}.
\bibitem[{Manhart et~al.(2008)Manhart, Peller and Brun}]{Manhart2008nutmodel}
\bibinfo{author}{Manhart, M.}, \bibinfo{author}{Peller, N.}, \bibinfo{author}{Brun, C.}, \bibinfo{year}{2008}.
\newblock \bibinfo{title}{Near-wall scaling for turbulent boundary layers with adverse pressure gradient}.
\newblock \bibinfo{journal}{Theor.~Comput.~Fluid~Dyn.} \bibinfo{volume}{22}, \bibinfo{pages}{243–260}.
\bibitem[{Morse and Mahesh(2021)}]{morse2021suboff}
\bibinfo{author}{Morse, N.}, \bibinfo{author}{Mahesh, K.}, \bibinfo{year}{2021}.
\newblock \bibinfo{title}{\text{Large-eddy} simulation and streamline coordinate analysis of flow over an axisymmetric hull}.
\newblock \bibinfo{journal}{J.~Fluid~Mech.} \bibinfo{volume}{926}, \bibinfo{pages}{A18}.
\bibitem[{Nicoud and Ducros(1999)}]{Nicoud1999WALE}
\bibinfo{author}{Nicoud, F.}, \bibinfo{author}{Ducros, F.}, \bibinfo{year}{1999}.
\newblock \bibinfo{title}{Subgrid-scale stress modelling based on the square of the velocity gradient tensor}.
\newblock \bibinfo{journal}{Flow~Turbul.~Combust.} \bibinfo{volume}{62}, \bibinfo{pages}{183--200}.
\bibitem[{{Ortiz-Tarin} et~al.(2021){Ortiz-Tarin}, Nidhan,  and Sarkar}]{Ortiz2021slenderBody}
\bibinfo{author}{{Ortiz-Tarin}, J.}, \bibinfo{author}{Nidhan, S.}, , \bibinfo{author}{Sarkar, S.}, \bibinfo{year}{2021}.
\newblock \bibinfo{title}{\text{{High-Reynolds-number}} wake of a slender body}.
\newblock \bibinfo{journal}{J.~Fluid~Mech.} \bibinfo{volume}{918}, \bibinfo{pages}{A30}.
\bibitem[{Owen et~al.(2020)Owen, Chrysokentis, Avila, Mira, Houzeaux, Borrell, Cajas and Lehmkuhl}]{Owen2020WMLES}
\bibinfo{author}{Owen, H.}, \bibinfo{author}{Chrysokentis, G.}, \bibinfo{author}{Avila, M.}, \bibinfo{author}{Mira, D.}, \bibinfo{author}{Houzeaux, G.}, \bibinfo{author}{Borrell, R.}, \bibinfo{author}{Cajas, J.}, \bibinfo{author}{Lehmkuhl, O.}, \bibinfo{year}{2020}.
\newblock \bibinfo{title}{\text{Wall-modeled} large-eddy simulation in a finite element framework}.
\newblock \bibinfo{journal}{Intl~J.~Numer.~Meth.~Fluids} \bibinfo{volume}{92}, \bibinfo{pages}{20--37}.
\bibitem[{Park and Moin(2016)}]{Park2016wallPressure}
\bibinfo{author}{Park, G.I.}, \bibinfo{author}{Moin, P.}, \bibinfo{year}{2016}.
\newblock \bibinfo{title}{{Space-time characteristics of wall-pressure and wall shear-stress fluctuations in wall-modeled large eddy simulation}}.
\newblock \bibinfo{journal}{Phys.~Rev.~Fluid} \bibinfo{volume}{1}, \bibinfo{pages}{024404}.
\bibitem[{Posa and Balaras(2016)}]{posa2016suboff}
\bibinfo{author}{Posa, A.}, \bibinfo{author}{Balaras, E.}, \bibinfo{year}{2016}.
\newblock \bibinfo{title}{\text{A} numerical investigation of the wake of an axisymmetric body with appendages}.
\newblock \bibinfo{journal}{J.~Fluid~Mech.} \bibinfo{volume}{792}, \bibinfo{pages}{470--498}.
\bibitem[{Posa and Balaras(2020)}]{posa2020suboff}
\bibinfo{author}{Posa, A.}, \bibinfo{author}{Balaras, E.}, \bibinfo{year}{2020}.
\newblock \bibinfo{title}{\text{A} numerical investigation about the effects of {Reynolds} number on the flow around an appended axisymmetric body of revolution}.
\newblock \bibinfo{journal}{J.~Fluid~Mech.} \bibinfo{volume}{884}, \bibinfo{pages}{A41}.
\bibitem[{Posa et~al.(2023)Posa, Felli and Broglia}]{Posa2023propeller}
\bibinfo{author}{Posa, A.}, \bibinfo{author}{Felli, M.}, \bibinfo{author}{Broglia, R.}, \bibinfo{year}{2023}.
\newblock \bibinfo{title}{{Acoustic far field of a propeller working in the wake of a hydrofoil}}.
\newblock \bibinfo{journal}{Phys.~Fluids} \bibinfo{volume}{35}, \bibinfo{pages}{125121}.
\bibitem[{Qu et~al.(2021)Qu, Wu, Zhao, Huang, Fu and Wang}]{Qu2021suboff}
\bibinfo{author}{Qu, Y.}, \bibinfo{author}{Wu, Q.}, \bibinfo{author}{Zhao, X.}, \bibinfo{author}{Huang, B.}, \bibinfo{author}{Fu, X.}, \bibinfo{author}{Wang, G.}, \bibinfo{year}{2021}.
\newblock \bibinfo{title}{\text{Numerical} investigation of flow structures around the {DARPA SUBOFF} model}.
\newblock \bibinfo{journal}{Ocean Eng.} \bibinfo{volume}{239}, \bibinfo{pages}{109866}.
\bibitem[{Rocca et~al.(2022)Rocca, Cianferra, Broglia and Armenio}]{Rocca2022BB2}
\bibinfo{author}{Rocca, A.}, \bibinfo{author}{Cianferra, M.}, \bibinfo{author}{Broglia, R.}, \bibinfo{author}{Armenio, V.}, \bibinfo{year}{2022}.
\newblock \bibinfo{title}{\text{Computational} hydroacoustic analysis of the {BB2} submarine using the advective {Ffowcs Williams and Hawkings} equation with {Wall-Modeled LES}}.
\newblock \bibinfo{journal}{Appl.~Ocean Res.} \bibinfo{volume}{129}, \bibinfo{pages}{103360}.
\bibitem[{Tang and Yu(2019)}]{Tang2019Noisebook}
\bibinfo{author}{Tang, W.L.}, \bibinfo{author}{Yu, M.S.}, \bibinfo{year}{2019}.
\newblock \bibinfo{title}{\text{Theory} of Hydrodynamic Noise}.
\newblock \bibinfo{publisher}{Science Press}.
\bibitem[{Wang et~al.(2006)Wang, Freund and Lele}]{Wang2006noiseReview}
\bibinfo{author}{Wang, M.}, \bibinfo{author}{Freund, J.B.}, \bibinfo{author}{Lele, S.K.}, \bibinfo{year}{2006}.
\newblock \bibinfo{title}{\text{Computational} prediction of flow-generated sound}.
\newblock \bibinfo{journal}{Annu.~Rev.~Fluid Mech.} \bibinfo{volume}{38}, \bibinfo{pages}{483--512}.
\bibitem[{Wang et~al.(2021)Wang, Huang and Pan}]{wang2021numerical}
\bibinfo{author}{Wang, X.}, \bibinfo{author}{Huang, Q.}, \bibinfo{author}{Pan, G.}, \bibinfo{year}{2021}.
\newblock \bibinfo{title}{\text{Numerical} research on the influence of sail leading edge shapes on the hydrodynamic noise of a submarine}.
\newblock \bibinfo{journal}{Appl.~Ocean Res.} \bibinfo{volume}{117}, \bibinfo{pages}{102935}.
\bibitem[{Xie et~al.(2019)Xie, Deng and Liao}]{Xie2019High-fidelitysolver}
\bibinfo{author}{Xie, B.}, \bibinfo{author}{Deng, X.}, \bibinfo{author}{Liao, S.J.}, \bibinfo{year}{2019}.
\newblock \bibinfo{title}{\text{High-fidelity} solver on polyhedral unstructured grids for low-{Mach} number compressible viscous flow}.
\newblock \bibinfo{journal}{Comput.~Methods Appl.~Mech.~Eng.} \bibinfo{volume}{357}.
\bibitem[{Xie et~al.(2020)Xie, Jin, Du and Liao}]{Xie2020consistent}
\bibinfo{author}{Xie, B.}, \bibinfo{author}{Jin, P.}, \bibinfo{author}{Du, Y.P.}, \bibinfo{author}{Liao, S.J.}, \bibinfo{year}{2020}.
\newblock \bibinfo{title}{\text{A} consistent and balanced-force model for incompressible multiphase flows on polyhedral unstructured grids}.
\newblock \bibinfo{journal}{Int.~J.~Multiph.~Flow} \bibinfo{volume}{122}, \bibinfo{pages}{103125}.
\bibitem[{Yang et~al.(2017)Yang, Park and Moin}]{Yang2017WMLES}
\bibinfo{author}{Yang, X.}, \bibinfo{author}{Park, G.}, \bibinfo{author}{Moin, P.}, \bibinfo{year}{2017}.
\newblock \bibinfo{title}{\text{Log-layer} mismatch and modeling of the fluctuating wall stress in wall-modeled large-eddy simulations}.
\newblock \bibinfo{journal}{Phys.~Rev.~Fluids} \bibinfo{volume}{2}, \bibinfo{pages}{104601}.
\bibitem[{Yu et~al.(2007)Yu, Wu and Pang}]{Yu2007shipNoise}
\bibinfo{author}{Yu, M.S.}, \bibinfo{author}{Wu, Y.S.}, \bibinfo{author}{Pang, Y.Z.}, \bibinfo{year}{2007}.
\newblock \bibinfo{title}{\text{A} review of progress for hydrodynamic noise of ships}.
\newblock \bibinfo{journal}{J.~Ship Mech.} \bibinfo{volume}{11}, \bibinfo{pages}{152--158}.
\newblock \bibinfo{note}{(in Chinese)}.
\bibitem[{Zhou et~al.(2022)Zhou, Xu, Wang and He}]{Zhou2022suboff}
\bibinfo{author}{Zhou, Z.T.}, \bibinfo{author}{Xu, Z.Y.}, \bibinfo{author}{Wang, S.Z.}, \bibinfo{author}{He, G.W.}, \bibinfo{year}{2022}.
\newblock \bibinfo{title}{{Wall-modeled large-eddy simulation of noise generated by turbulence around an appended axisymmetric body of revolution}}.
\newblock \bibinfo{journal}{J.~Hydrodynam.~B} \bibinfo{volume}{34}, \bibinfo{pages}{533--554}.
\bibitem[{Özden et~al.(2019)Özden, Özden, Demir and Kuroğlu}]{Ozden2019}
\bibinfo{author}{Özden, Y.A.}, \bibinfo{author}{Özden, M.C.}, \bibinfo{author}{Demir, E.}, \bibinfo{author}{Kuroğlu, S.}, \bibinfo{year}{2019}.
\newblock \bibinfo{title}{\text{Experimental} and numerical investigation of {DARPA Suboff} submarine propelled with {INSEAN E1619} propeller for self-propulsion}.
\newblock \bibinfo{journal}{J.~Ship~Res.} \bibinfo{volume}{63}, \bibinfo{pages}{235--250}.

\end{thebibliography}


\end{document}